\documentclass{aa}

\usepackage{graphicx,txfonts,color}
\usepackage{longtable,lscape}
\usepackage{multirow,multicol}
\usepackage{natbib}
\usepackage{xspace}
\usepackage{amsmath}
\usepackage{fixltx2e}

\newcommand{\water}{H$_2$O\xspace}
\newcommand{\HH}{H$_2$\xspace}

\newcommand{\Msunyr}{$M_{\odot}$\,yr$^{-1}$\xspace}

\newcommand{\vterminal}{$\upsilon_{\infty}$\xspace}
\newcommand{\kms}{km\,s$^{-1}$\xspace}
\newcommand{\Rnk}{$R_{\rm{NK}}$\xspace}
\newcommand{\Eup}{$E_{\rm{up}}$\xspace}
\newcommand{\chired}{$\chi^2_{red}$\xspace}
\newcommand{\vlsr}{$\varv_{\rm{LSR}}$\xspace}

\begin{document}
   \title{A HIFI view on circumstellar H$_2$O in M-type AGB stars:\\
   radiative transfer, velocity profiles, and H$_2$O line cooling}


   \author{M. Maercker \inst{1}
          \and
          T. Danilovich \inst{1}
    	\and
	H. Olofsson \inst{1}
          \and
          E. De Beck \inst{1}
          \and
          K. Justtanont \inst{1}
          \and
          R. Lombaert \inst{1}
         \and
         P. Royer \inst{2}
          }

   \offprints{M. Maercker}

   \institute{Onsala Space Observatory, Dept. of Radio and Space Science, Chalmers University of Technology, SE-43992 Onsala, Sweden\\
   \email{maercker@chalmers.se}
   \and
   Instituut voor Sterrenkunde, KU Leuven, Celestijnenlaan 200D 2401, 3001 Leuven, Belgium\\}

   \date{}
   
   
  \abstract
  {}
   {We aim to constrain the temperature and velocity structures, and \water abundances in the winds of a sample of M-type Asymptotic Giant Branch (AGB) stars. We further aim to determine the effect of \water line cooling on the energy balance in the inner circumstellar envelope.}
   {We use two radiative-transfer codes to model molecular emission lines of CO and \water towards four M-type AGB stars. We focus on spectrally resolved observations of CO and \water from HIFI aboard the \emph{Herschel Space Observatory}. The observations are complemented by ground-based CO observations, and spectrally unresolved CO and \water observations with PACS aboard \emph{Herschel}. The observed line profiles constrain the velocity structure throughout the circumstellar envelopes (CSEs), while the CO intensities constrain the temperature structure in the CSEs. The \water observations constrain the o-\water and p-\water abundances relative to H$_2$. Finally, the radiative-transfer modelling allows to solve the energy balance in the CSE, in principle including also \water line cooling.}
   {The fits to the line profiles only set moderate constraints on the velocity profile, indicating shallower acceleration profiles in the winds of M-type AGB stars than predicted by dynamical models, while the CO observations effectively constrain the temperature structure. Including \water line cooling in the energy balance was only possible for the low-mass-loss-rate objects in the sample, and required an ad hoc adjustment of the dust velocity profile in order to counteract extreme cooling in the inner CSE. \water line cooling was therefore excluded from the models. The constraints set on the temperature profile by the CO lines nevertheless allowed us to derive \water abundances. The derived \water abundances confirm previous estimates and are consistent with chemical models. However, the uncertainties in the derived abundances are relatively large, in particular for p-\water, and consequently the derived o/p-\water ratios are not well constrained.}
{}

    \keywords{Stars: AGB and post-AGB -  Stars: evolution - Stars: late-type - Stars: mass loss - Stars: abundances 
               }
\titlerunning{A HIFI view on circumstellar H$_2$O in M-type AGB stars}
   \maketitle

\section{Introduction}
\label{s:intro}
AGB stars play an important role in the chemical evolution of galaxies and the universe~\citep{bussoetal1999,schneideretal2014}. Elements created inside the star get returned to interstellar space in the form of molecules and dust grains in a wind from the stellar surface. This mass loss is driven by radiation pressure on dust grains formed in the upper atmosphere, which drag the gas along through collisions~\citep{bowen1988,hofneretal1998,simisetal2001,hofner2008}. The lifetime on the AGB is determined through the mass loss. 

In order to determine stellar elemental yields to interstellar space from AGB stars, and to understand their role in the chemical evolution of the universe, it is important to determine the molecular content of the CSE, the chemical reaction paths, and the characteristics of the mass-loss process. Observations of molecular emission lines have been used to determine these properties. In particular, observations of CO emission lines have proven to be a useful tool to describe the mass loss~\citep[e.g.,][]{schoieretal2002,delgadoetal2003b,ramstedtetal2008,debecketal2010,khourietal2014a}. Observations of other molecules have led to important insights in the chemistry and structure of AGB CSEs~\citep[e.g.,][]{delgadoetal2003b,schoieretal2007,maerckeretal2008,maerckeretal2009,decinetal2010b,cernicharoetal2010,cernicharoetal2011,debecketal2012,justtanontetal2012,khourietal2014b,lombaertetal2016}.

A molecule of particular importance for M-type AGB stars is \water. It is one of the most abundant molecules in the winds of these stars, rivalled only by CO and H$_2$ \citep{cherchneff2006,maerckeretal2008,maerckeretal2009,decinetal2010b,decinetal2010c,khourietal2014b}. As an oxygen-bearing molecule it plays an important role in the oxygen chemistry  and the formation of other molecules. Due to its many far infrared transitions it is a dominant coolant in the inner CSE~\citep{truong-bachetal1999}. The \water lines are emitted in the warm, inner envelope, and observations of emission lines also probe the acceleration of the wind~\citep{decinetal2010b,decinetal2010c}. As such, observations of \water serve as an excellent diagnostic for the chemistry and physics of AGB CSEs. At the same time, a correct treatment of \water and its effects on the physical structure of the CSE is essential in all radiative-transfer calculations and mass-loss descriptions.

Observations of \water lines are difficult to obtain from ground-based telescopes due to the absorption in Earth's atmosphere~\citep[although it is possible to observe masers and vibrationally excited lines in the $\nu_2=1$ bending mode;][]{mentenetal2006}. The \emph{Infrared Space Observatory (ISO)} observed a large number of \water lines in the wavelength range between 43 $\mu$m and 197 $\mu$m towards AGB stars~\citep[e.g.,][]{barlowetal1996,truong-bachetal1999,maerckeretal2008}. Radiative-transfer models generally require relatively high amounts of \water to fit the observations of \water emission lines towards the CSEs of M-type AGB stars. However, the ISO observations are spectrally unresolved, increasing the uncertainty in the determined \water abundances, and entirely eliminating any information on the velocity field of the \water line emitting region. The \emph{SWAS}~\citep{melnicketal2000} and \emph{Odin}~\citep{nordhetal2003} satellites provided spectrally resolved observations of the ground-state ortho-\water($1_{10}-1_{01}$) transition at 557 GHz towards M-type AGB stars. Modelling the line-width of this transition further constrained the determined \water abundances, and demonstrated that the \water line emitting region extends into the wind acceleration zone~\citep{justtanontetal2005,hasegawaetal2006,maerckeretal2009}.

The Heterodyne Instrument for the Far-infrared (\emph{HIFI}) aboard the Herschel Space Observatory (\emph{Herschel}) for the first time provided high-sensitivity and spectrally resolved observations of a large number of \water lines towards AGB CSEs. The observations used here were done as part of the HIFISTARS guaranteed time key program (P.I.: V. Bujarrabal) to study the CSEs around evolved stars of all chemical types, mainly targeting the CO and \water lines (see Sect.~\ref{s:hifiobs}). In particular, HIFISTARS contained a sample of M-type AGB stars for which, in addition to CO and \water, a total of seven different molecules were detected~\citep{justtanontetal2012}. The HIFI observations allow us to determine the abundance and distribution of \water in the CSEs of the observed sources, as well as to constrain the velocity profile, through radiative-transfer modelling of the observed CO and \water lines.

In this paper we use advanced radiative-transfer models in order to determine the abundance of \water in the CSEs of M-type AGB stars, constrain the velocity profile of the stellar wind, as well as explore the significance of \water line cooling in the energy balance of  the inner CSE. The basic modelling method has been used previously for models of \water lines~\citep{maerckeretal2008,maerckeretal2009}, and has been upgraded to also include velocity profiles and the effect of \water line cooling on the energy balance~\citep{schoieretal2011,danilovichetal2014}. In addition to the HIFI observations, we use spectrally unresolved lines of CO and \water observed with the Photoconductor Array Camera and Spectrometer (\emph{PACS}) aboard \emph{Herschel} to further constrain the models. All abundances are relative to H$_2$. In Sect.~\ref{s:obs} we present the observations and basic information on the modelled M-type AGB stars. In Sect.~\ref{s:radtransf} the radiative-transfer models are presented. We present and discuss the results in Sects.~\ref{s:results} and~\ref{s:discussion}, respectively, and summarise our conclusions in Sect.~\ref{s:conclusions}.

\section{Observations}
\label{s:obs}

We model rotational lines from the main isotopologues  of CO and \water from both ground-based and space telescopes. The observations have been published previously, and are summarised below.

\subsection{Ground-based observations of CO}
\label{s:groundCO}

We included ground-based observations in the molecular line radiative modelling of CO. The observations were taken at Onsala Space Observatory, the JCMT, and SEST~\citep[for details, see][and Table~\ref{t:COlines}]{kerschbaumco1999,delgadoetal2003b,justtanontetal2005,ramstedtetal2008,maerckeretal2008}. Transitions from CO($1-0$) to CO($4-3$) were included for all sources except R~Dor, which lacks CO($4-3$). The uncertainty in the integrated intensities is assumed to be 20\%.

\subsection{HIFI observations of CO and \water}
\label{s:hifiobs}

The HIFI observations covered a number of expected strong CO and \water lines in the frequency range 556.6--1843.5 GHz. For CO the CO($6-5$), CO($10-9$), and CO($16-15$) were observed. The observed integrated intensities for CO and the observed \water lines are presented in Tables~\ref{t:COlines} to~\ref{t:pH2Olines}. These observations were already presented in \cite{justtanontetal2012}, but were reprocessed using the latest main beam efficiencies. For some of the bands this led to an increase in the integrated fluxes of up to 20\%. We assume an average uncertainty in the integrated intensity for all lines of 20\%.

\subsection{PACS observations of CO and \water}
\label{s:pacsobs}

The PACS observations were taken as part of the MESS guaranteed-time key project \citep{groenewegenetal2011}, which covered a sample of AGB stars of all chemical types. The observations and the detailed data reduction will be presented in a forthcoming paper dedicated to the M-type AGB stars of the MESS sample (Decin et al.~in prep).  The PACS spectra cover the entire frequency range from 1580 up to 5450 GHz. For this study, we only include non-blended CO and H2O emission lines. For a detailed description on how line blends are identified, we refer to~\cite{lombaertetal2016}. The observed integrated intensities are presented in Tables~\ref{t:COlines} to~\ref{t:pH2Olines}. We assume an average uncertainty in the integrated intensity for all PACS lines of 20


\subsection{Sources}
\label{s:sample}

The HIFISTARS program included nine M-type AGB stars. Of these we include only fours stars: R~Dor, R~Cas, TX~Cam, and IK~Tau. R~Dor is a SRb variable, and the remaining sources are Mira-type variables. The HIFI observations of IK~Tau were modelled by \cite{decinetal2010c} using a different radiative-transfer code. We include IK~Tau here for a comparison of the modelling approaches. The HIFI observations of W~Hya were modelled in detail by \cite{khourietal2014a,khourietal2014b} using the same code as \cite{decinetal2010c}. Mira has a very complex circumstellar environment due to binary interaction~\citep{ramstedtetal2014}, and the remaining sources are OH/IR stars with high mass-loss rates. These sources are therefore not included in this study. 

The mass-loss rates and \water abundances of the remaining sources have been modelled previously~\citep{maerckeretal2008,maerckeretal2009}. The lowest and highest mass-loss rates of the sample were estimated to be $2\times10^{-7}$\,\Msunyr (R~Dor) and $1\times10^{-5}$\,\Msunyr (IK~Tau), respectively, based on ground-based CO observations only. Ortho-\water abundances were estimated to be between $2.0\times10^{-4}$ (R~Dor) and $3.5\times10^{-4}$ (IK~Tau) relative to H$_2$, based on ISO observations. For the Mira-variables  R~Cas, TX~Cam and IK~Tau the luminosities were derived from a period-luminosity relation~\citep{feastetal1989} and the distances were derived in the dust modelling, while for the SRb variable R~Dor we assumed a distance of 59\,pc and derived the luminosity in the dust modelling. The same procedure was used in~\cite{maerckeretal2008}.

\section{Radiative-transfer modelling}
\label{s:radtransf}

The radiative-transfer modelling is done by combining results from models of the dust radiation field, CO radiative-transfer models, and \water radiative-transfer models. All models assume a spherically symmetric, homogenous wind produced by a constant mass-loss rate. The procedure allows us to determine the dust temperature profile, the mass-loss rate, the kinetic temperature profile, the gas expansion velocity profile, and the fractional \water abundance (relative to \HH) distribution throughout the CSE. Since the aim of the paper is to compare the derived \water abundances between the sources, derive gas-velocity profiles, and investigate the significance of \water line cooling, we treat the sources as homogenously as possibly. Therefore basic properties, such as the dust grain sizes, dust composition, and the initial wind velocity, are assumed to be the same for all sources. Although detailed modelling of these parameters for individual sources may result in different values, the uncertainties in these individual models would complicate the comparative study that is the goal of this work. This basic modelling strategy has been thoroughly tested and used previously \citep[e.g.,][]{schoierco2001,olofssonetal2002,ramstedtetal2008,maerckeretal2008,maerckeretal2009,schoieretal2011,danilovichetal2014}. 

\subsection{The basic CO model}
\label{s:basicco}

The code used for the modelling of the CO lines is described in detail in \cite{schoierco2001}. It employs the Monte Carlo method and is a non-local, non-LTE radiative-transfer code that solves the energy balance and derives a self-consistent kinetic temperature profile. We include radiation from dust by calculating the spectral energy distribution using DUSTY~\citep{ivezicetal1999}, constrained by 2MASS and IRAS fluxes. We used amorphous silicate dust with optical constants from \cite{justtanontco1992}, assuming spherical grains with a radius of 0.05 $\mu$m and a density of 3 g\,cm$^{-3}$. The dust model calculates the dust condensation radius, the dust optical depth, $\tau_{\rm{d}}$, at 10\,$\mu$m and gives a dust temperature profile, both of which are included in the CO model. Our dust models are largely based on the results presented in \cite{maerckeretal2008}, with possible slight adjustments in the luminosity and/or dust optical depth. The radius of the CO envelope is determined by models of photodissociation of CO~\citep{mamonetal1988}. The code and modelling method has been used extensively to model the emission lines from various molecules, including CO~\citep[e.g.,][]{olofssonetal2002}, SiO~\citep{ramstedtetal2009}, and HCN~\citep{schoieretal2013}. 

Models of CO lines up to the $J=4-3$ transition were used to determine the basic properties of the CSE in the modelling of \water lines observed by ISO and Odin~\citep{maerckeretal2008,maerckeretal2009}. The emission of these lines is dominated by the outer parts of the envelope, where the wind has already reached its terminal velocity. HIFI provides new observations of higher-energy CO transitions. These are emitted in the inner CSE, and it is now possible to constrain the velocity profile. The models describe a velocity law for the gas and the dust separately. The gas-velocity profile follows the form 

\begin{equation}
\label{e:vprof}
\upsilon_g(r)=\upsilon_{g,i}+(\upsilon_{g,\infty}-\upsilon_{g,i})\times\left(1-\frac{{R_i}} {{r}}\right)^{\beta},
\end{equation} 

\noindent
where $\upsilon_{\rm{g,i}}$ is the gas expansion velocity at the inner radius (assumed to be 3 \kms in all models), $\upsilon_{g,\infty}$ is the gas terminal velocity, $R_{\rm{i}}$ is the inner radius (set to the dust condensation radius), and $\beta$ describes how fast the wind accelerates. The dust expansion velocity is connected to the gas expansion through the dust drift velocity $\upsilon_{dr}(r)$:

\begin{equation}
\label{e:vdust}
\upsilon_d(r)=\upsilon_g(r)+\upsilon_{dr}(r),
\end{equation}

\noindent
where $\upsilon_{dr}(r)$ follows the same shape as Eq.~\ref{e:vprof}, with $\upsilon_{\rm{dr,i}}$ the drift velocity at the inner radius. The terminal drift velocity $\upsilon_{dr,\infty}$ is calculated by

\begin{equation}
\label{e:vdterm}
\upsilon_{dr,\infty}=\sqrt{ \frac{L_{\odot} \upsilon_{g,\infty}Q} { \dot M c}},
\end{equation}

\noindent
where $L$ is the stellar luminosity, $\upsilon_{g,\infty}$ is the terminal gas expansion velocity, $Q$ is the scattering efficiency assumed to be 3\%, $\dot M$ is the mass-loss rate, and $c$ is the speed of light. The terminal dust velocity is then given by $\upsilon_{d,\infty}=\upsilon_{dr,\infty}+\upsilon_{g,\infty}$. Note that there are no constraints on the dust velocity profile, and we assume Eqs.~\ref{e:vdust} and ~\ref{e:vdterm} to give reasonable values. However, the exact shape of the drift velocity profile strongly affects the heating in the inner envelope (see Sect.~\ref{s:heatcool}).

The gas expansion velocity at the inner radius, $\upsilon_{\rm{g,i}}$, equals typical values of the sound-speed in the inner winds of AGB stars~\citep[e.g.,][]{decinetal2006}. The gas-velocity profile is constrained by fitting the widths of the CO lines at the zero-intensity level. Low-$J$ lines constrain the terminal velocity, while higher-$J$ lines are emitted from closer to the star and constrain the shape of the expansion profile. This is not affected very much by the adopted mass-loss rate, and the expansion velocity profile can be constrained by initially assuming reasonable parameters for the mass-loss rate. We base our initial values on previous models of CO lines for these sources assuming constant expansion velocities. Once the new expansion velocity profile is determined, we proceed to model the mass-loss rate and temperature profile of the CSE to include in the \water modelling. The best-fit CO model is determined by calculating a grid of models varying the mass-loss rate and dust-to-gas ratio. To find the best fit to the data, a $\chi^2$ analysis comparing the observed line intensities with the model intensities is then performed by minimising 

\begin{equation}
\label{e:chi2}
\chi^2=\sum{\left(\frac{I_{\rm{obs,i}}-I_{\rm{mod,i}}}{\sigma_{\rm{obs,i}}}\right)^2},
\end{equation}

\noindent
where $I_{\rm{obs,i}}$ is the integrated intensity of the observation $i$, $I_{\rm{mod,i}}$ is the modelled intensity, $\sigma_{\rm{obs,i}}$ is the uncertainty in the observation. An error of 20\% in the integrated intensities is assumed for all the CO lines.

\subsection{The basic \water  model}
\label{s:basich2o}

We use an accelerated lambda iteration (ALI) method to calculate the radiative transfer of \water lines. The model is the same as described in \cite{maerckeretal2008}, but now includes a velocity profile in the same form as Eq.~\ref{e:vprof}. We include the same dust-radiation field and velocity profiles as for CO. The mass-loss rate and kinetic temperature profiles are included from the results of the CO modelling. The remaining free parameters in the \water models are the fractional \water abundance at the inner radius and the $e$-folding radius of the \water envelope~\citep[see][]{maerckeretal2008}. Together these describe the fractional abundance distribution of \water. We model both ortho- and para-\water. The best-fit \water model is found by performing the same $\chi^2$ analysis as for CO, but now on a grid of models in \water-abundance and radius of the \water envelope. An error of 20\% is assumed for the integrated intensities of the HIFI  and PACS lines. A good estimate of the \water radius is obtained from photodissociation models as the radius where the OH abundance peaks~\citep[as a photodissociation product of \water;][]{netzerco1987,maerckeretal2008,maerckeretal2009}. From here on, we will refer to the photodissociation radius of \water as \Rnk, given by

\begin{equation}
\label{e:rnk}
$\Rnk$=5.4\times10^{16} \left(\cfrac{\dot{M}}{10^{-5}}\right)^{0.7}\left(\cfrac{\upsilon_{g,\infty}}{\rm{km\,s^{-1}}} \right)^{-0.4}\,\rm{cm}.
\end{equation}

\subsection{Heating and cooling in the energy balance: the problem of \water line cooling}
\label{s:heatcool}

The CO model self-consistently solves the energy balance throughout the CSE by including all (known) heating and cooling terms~\citep{schoierco2001}. The effect of heating due to dust-gas collisions, photoelectric heating, and heat exchange between the dust and the gas is included. The total cooling includes cooling due to adiabatic expansion, and molecular line cooling from CO and H$_2$. In order to determine the effect of \water line cooling on the energy balance in the inner CSE, the contribution to the cooling by \water lines is given as output in the \water model. This can be included in the energy balance when calculating the CO model. 

To fully include \water line cooling, a careful iterative procedure is necessary, slowly increasing the contribution from \water line cooling in each step. Immediately including full \water line cooling typically leads to over-cooling in the inner envelope and results in negative temperatures. This iterative procedure was followed previously for the S-type stars $\chi$~Cyg and W~Aql using HIFI data as constraints~\citep[][respectively]{schoieretal2011,danilovichetal2014}. 

In the case of M-type AGB stars, the abundance of \water is high enough to make it one of the dominant sources of cooling, in particular in the inner envelope. Here we attempt to include \water line cooling in an iterative procedure for the M-type stars R~Dor and R~Cas. We start from the best-fit CO model that does not include \water line cooling. The kinetic temperature profile, velocity profile, and mass-loss rate are included as input in the \water models. The abundance of \water is then adjusted to fit the observed HIFI and PACS lines, and the \water line cooling is calculated (see Sect.~\ref{s:h2ocool} for details). Only a fraction of the resulting \water line cooling is then included in the CO model. This allows us to adjust the CO model in a way that counteracts extreme cooling and to ensure that the model still fits the observations by, e.g., increasing the mass-loss rate, increasing/decreasing the dust-to-gas ratio, and adjusting the drift-velocity profile. Once a new satisfactory CO model has been calculated, a new \water model is calculated based on the new parameters, giving a new \water abundance and hence cooling function. This procedure is repeated, increasing the fraction with which \water line cooling is included with every step. Full \water line cooling is included when 100\% of the cooling function is included in the energy balance, good fits have been obtained to the observed lines for both CO and \water, and the temperature profile has converged between iterations.

While this procedure was successful for the S-type stars, we come to the conclusion that it is not feasible for the high \water abundances in M-type AGB stars. The iterative procedure makes the modelling very computationally expensive. It is therefore unfeasible to run a grid of models to properly probe parameter space, making the model results very unreliable. For the same reason the \water $e$-folding radius is treated as a fixed parameter and is set to \Rnk. We nevertheless attempted to include 100\% \water line cooling for the low mass-loss-rate objects R~Dor and R~Cas and find that either a complete understanding of the heating and cooling processes in the CSE is lacking, or that the numerical implementation of radiative line cooling may have to be revised (see Sect.~\ref{s:h2ocool}). Although a sufficient number of CO-lines nevertheless constrain the temperature profile of the CSE, a correct treatment of \water line cooling is necessary for a full understanding of the physical processes in the CSEs of M-type AGB stars. Consequently the final CO and \water models do not include \water line cooling.

Note that~\cite{decinetal2010a} do manage to include \water line cooling in the energy balance for the M-type AGB star IK~Tau. However, they do not include any explicit discussion on the issues of \water line cooling, and it is not clear how they avoid the problems with the energy balance that we encounter in our modelling.

\section{Results}
\label{s:results}

\begin{figure*}[h]
\centering
\vspace{-0.5cm}
\includegraphics[width=19cm]{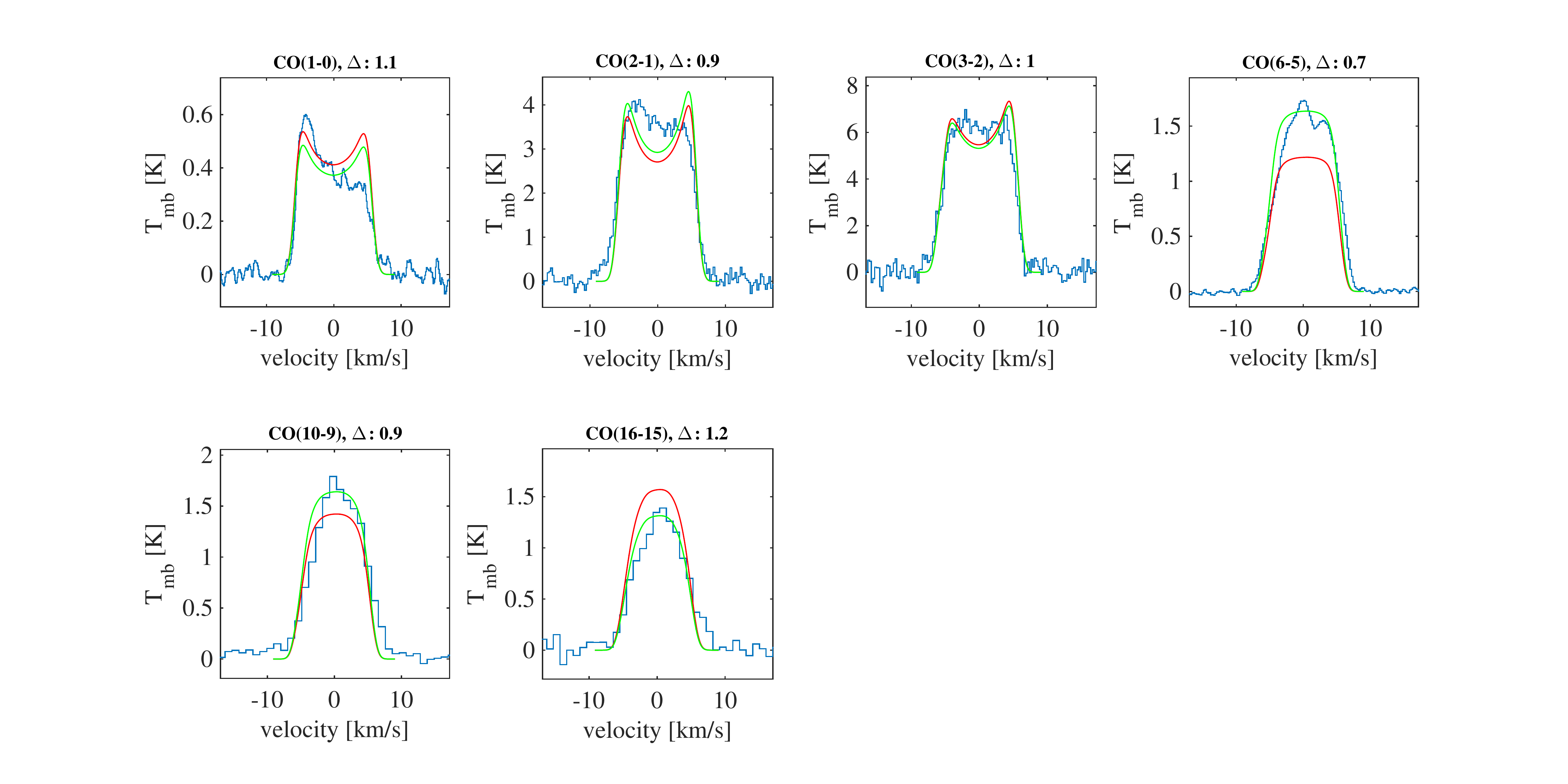}
\caption{Best-fit CO models for R~Dor. The blue histograms are the observations. The red lines are the model lines, the green lines are the model lines scaled to the same integrated intensities as the observations. The velocities are given with respect to the \vlsr= 6.5\,\kms.}
\label{f:colinesrdor}
\end{figure*}

\subsection{CO models}
\label{s:comods}

\begin{figure}
\centering
\includegraphics[width=9cm]{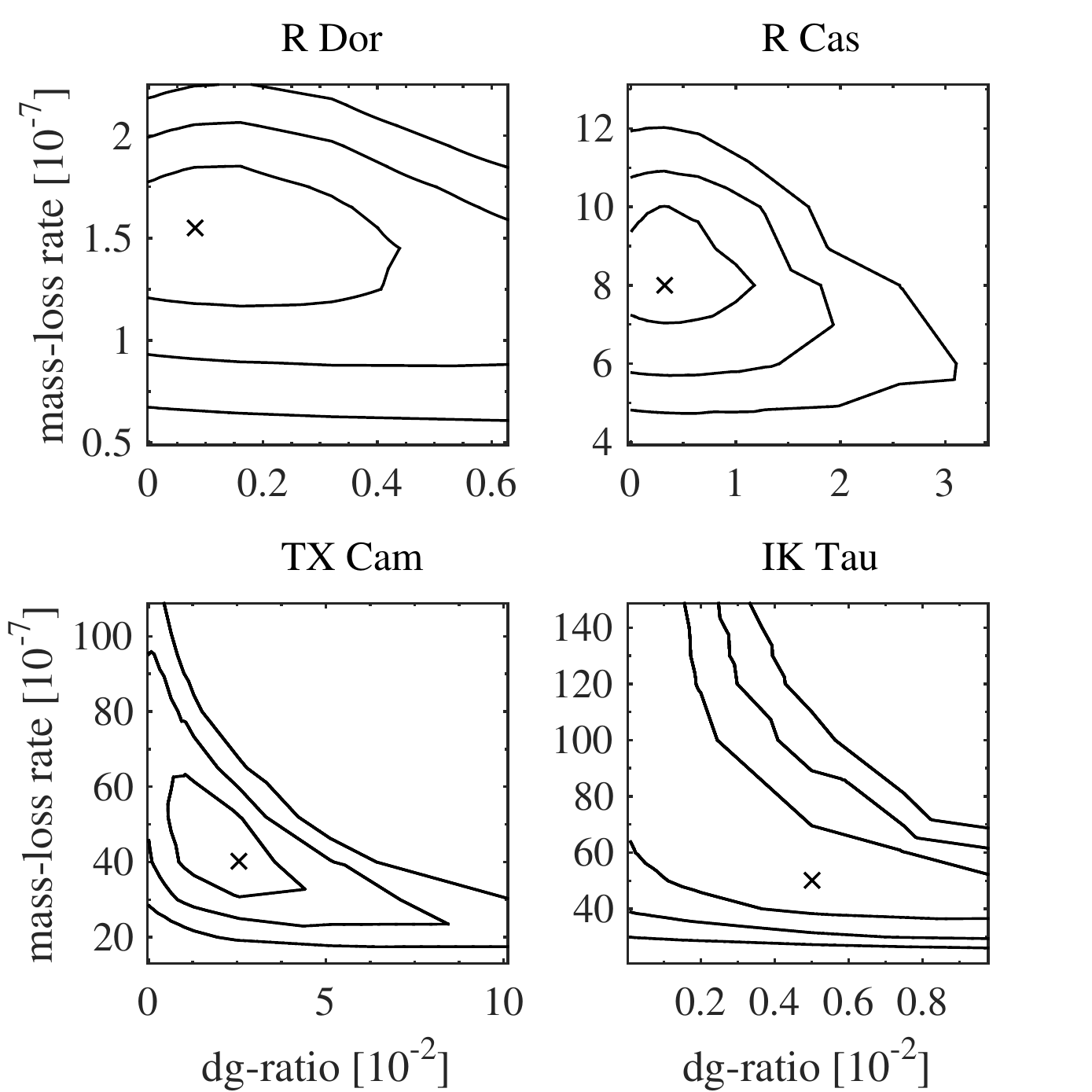}
\caption{$\chi^2$-maps of the CO model grids for R~Dor (top left), R~Cas (top right), TX~Cam (bottom left), and IK~Tau (bottom right). The contours show the 1$\sigma$, 2$\sigma$, and 3$\sigma$ levels. The cross marks the lowest $\chi^2_{red}$ model. }
\label{f:cochi2maps}
\end{figure}

\begin{figure}
\centering
\includegraphics[width=9cm]{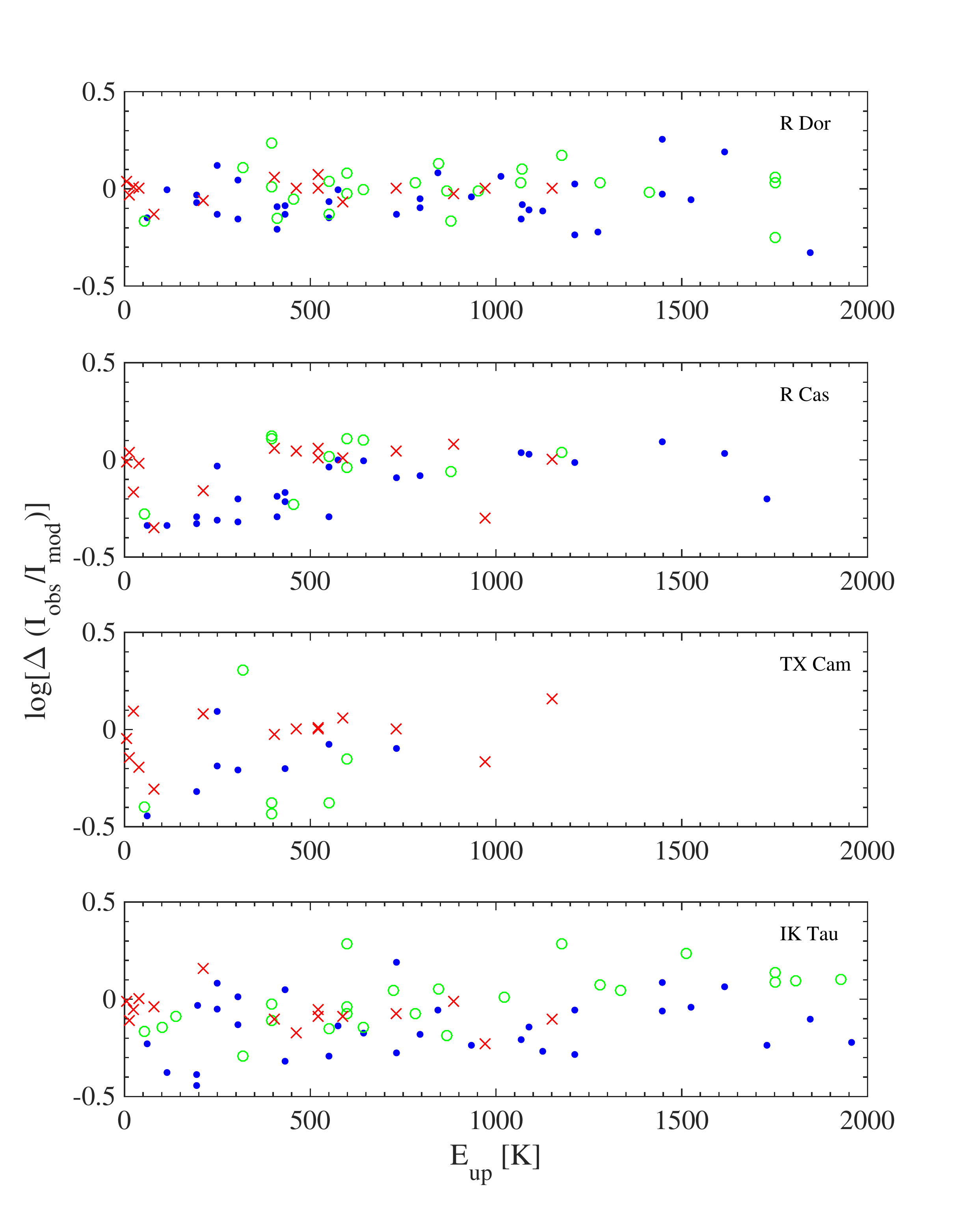}
\caption{Ratio between observed and modelled integrated intensities vs. upper-level energy for the different transitions (CO: red crosses, ortho-\water: blue dots, para-\water: green circles).}
\label{f:ratiovsenergy}
\end{figure}

Figure~\ref{f:colinesrdor} shows the best-fit CO models compared to the observed ground-based and HIFI lines for R~Dor. Figures~\ref{f:colinesrcas} to~\ref{f:colinesiktau} show the equivalent for R~Cas, TX~Cam, and IK~Tau, respectively. Table~\ref{t:COlines}  gives the observed and modelled integrated intensities of all CO lines, including the PACS lines, together with the upper-level energies (E$_{\rm{up}}$) and the ratios between models and observations for individual lines ($\Delta$). Fig.~\ref{f:cochi2maps} shows the $\chi^2_{red}$ maps for the different CO model grids and Table~\ref{t:coresults} gives the best-fit parameters. The integrated intensities of the individual observed lines are all reproduced to within $\approx$25\%. The average ratios between all observed and model lines ($\delta$) are all close to 1. Although observed by HIFI, the CO($16-15$) transition towards TX~Cam is extremely over-estimated in the model by an order of magnitude, while the PACS line is well reproduced. The HIFI line is therefore excluded in the $\chi^2$ analysis. The ratio between modelled and observed line intensities as a function of \Eup is shown in Fig.~\ref{f:ratiovsenergy} (red crosses). Generally the lines scatter around a ratio of one, although there might be a slight correlation, with models preferentially under-predicting the intensities for low-\Eup transitions.
 
For all models we use the photodissociation radius described by \cite{mamonetal1988}. This generally leads to overly resolved model line-profiles for the lower transitions, as evidenced by the double-peaked model line-profiles~\citep[specifically discussed for R~Dor by][]{olofssonetal2002}. This is a common problem in CO radiative-transfer models~\citep[e.g.,][]{justtanontetal2005,maerckeretal2008,danilovichetal2015}. We attempted to fit the low-$J$ line profiles by reducing the CO envelope size. However, this generally leads to an underestimation of the total integrated intensity compared to the high-$J$ transitions. A possible way of correcting for this is by changing the dust parameters and the velocity field. However, the angular size of the envelope also depends on the assumed and uncertain distance to the source. Since this is a degenerate problem, we chose to fix the radius to the CO photodissociation radius and use the integrated intensities when determining the best-fit models. 
 
\begin{figure}
\centering
\includegraphics[width=9cm]{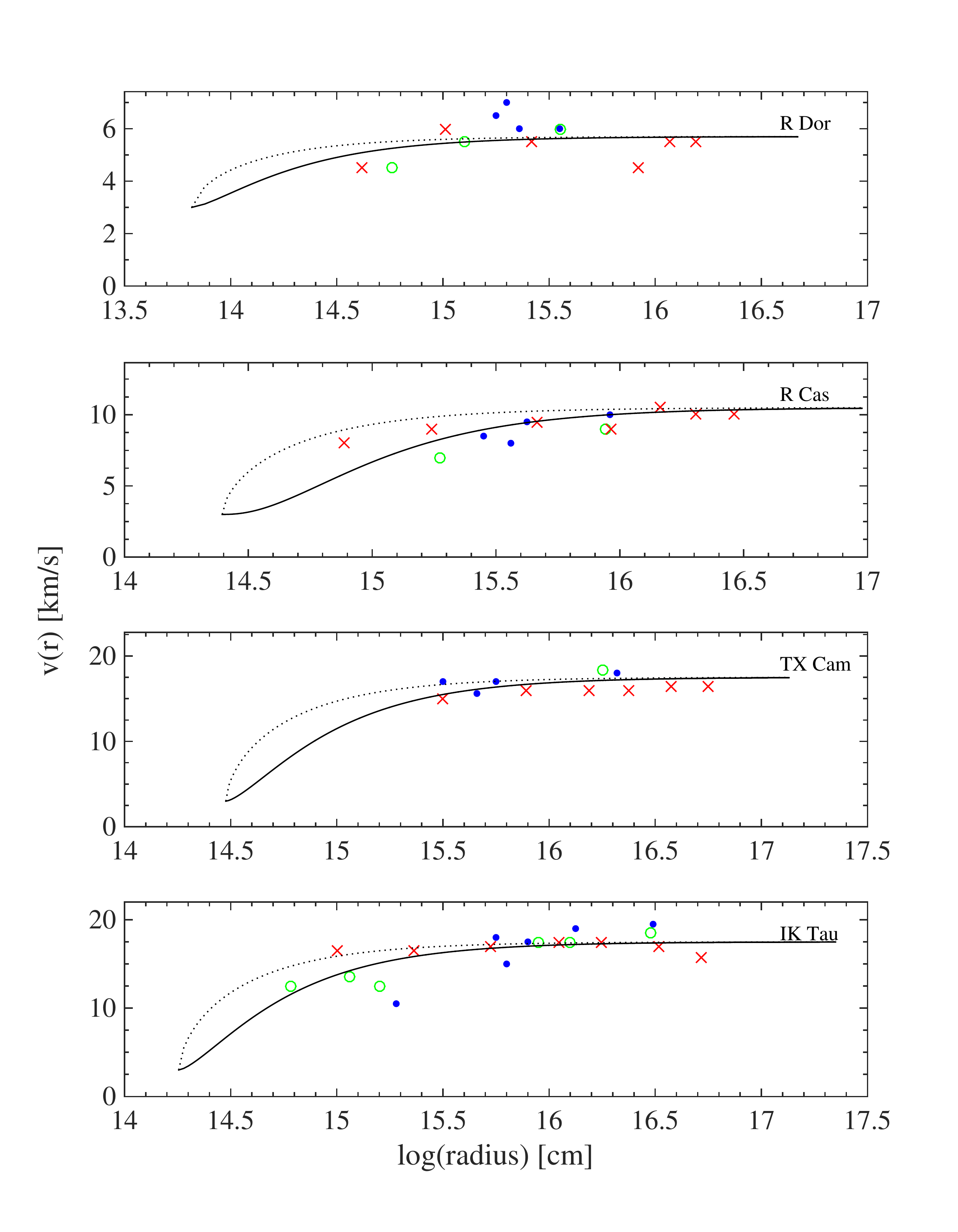}
\caption{Velocity profiles derived for R~Dor, R~Cas, TX~Cam and IK~Tau (top to bottom) compared to the measured line-widths from lines emitted at different radii in the CSE (CO: red crosses, ortho-\water: blue dots, para-\water: green circles. The widths are measured at zero power. The radii are determined by the peak in the brightness profiles of each transition). The solid black lines give the profiles with the $\beta$ given in Table~\ref{t:coresults}, the dotted black lines give the profile for $\beta=0.6$.}
\label{f:vprofiles}
\end{figure}

In principle the terminal expansion velocity of the gas, $\upsilon_{g,\infty}$, should be well constrained by the width at zero power (in the following referred to as the width of the line) of the low-$J$ CO transitions, while the widths of the higher-$J$ CO transitions and \water transitions constrain the shape of the velocity profile ($\beta$ in Eq.~\ref{e:vprof}). We attempt to constrain the velocity law using the observed CO and \water lines, adjusting $\beta$ in the radiative-transfer models to fit the observed line widths. For R~Cas we derive $\beta=2.5$, and for the remaining three sources $\beta$=1.5. For IK~Tau the momentum equation gives a steeper profile with $\beta=0.6$~\citep{decinetal2010c}. Although comparatively shallow velocity profile seem to be preferred, the $\beta$ is not very well constrained by the observed CO and \water lines (Fig.~\ref{f:vprofiles}), and we estimate the uncertainty in $\beta$ to be $\pm1$.

The estimated mass-loss rates are consistent with previous estimates using the same radiative-transfer code and based on ground-based data alone~\citep{maerckeretal2008}. Only for TX~Cam do we derive a mass-loss rate a factor of $\approx$2 higher than the previous estimate $-$ still within the absolute uncertainty of independent mass-loss rate estimates of a factor of three~\citep{ramstedtetal2008}. The uncertainty in the mass-loss rate within the adopted model is $\approx$30\%. For IK~Tau it proved difficult to find an upper limit to the mass-loss rate, likely due to a saturation effect of the CO lines at high mass-loss rates, caused by increasing line optical depths and cooler envelopes~\citep{ramstedtetal2008}. See Sect.~\ref{s:codecomp} for a comparison with similar codes. 

The derived dust-to-gas ratios have considerable uncertainties. For R~Dor and R~Cas we derive values of a few times $10^{-4}$ with relatively strong constraints on the upper limit. For TX~Cam we derive a value a factor $\approx$50 larger. For IK~Tau the best-fit value is $\approx$10 times larger than for R~Dor and R~Cas, and agrees with chemical models at 6\,R$_*$~\citep{gobrechtetal2016}, however no limits can be set. \cite{ramstedtetal2008} derive values of a few times $10^{-3}$. The dust-to-gas ratio mainly affects the energy balance, where it is part of the heating of the gas through dust-gas collisions. This heating term depends degenerately also on the grain size and density. For this reason~\cite{schoierco2001} chose to combine these parameters in the so-called \emph{h}-parameter. The derived d/g-ratios give \emph{h}-parameters of 0.05, 0.03, 1.7, and 0.3 for R~Dor, R~Cas, TX~Cam, and IK~Tau, respectively. \cite{ramstedtetal2008} derive values of 1.0 and 0.3 for TX~Cam and IK~Tau, respectively. 

\begin{figure}
\centering
\includegraphics[width=9cm]{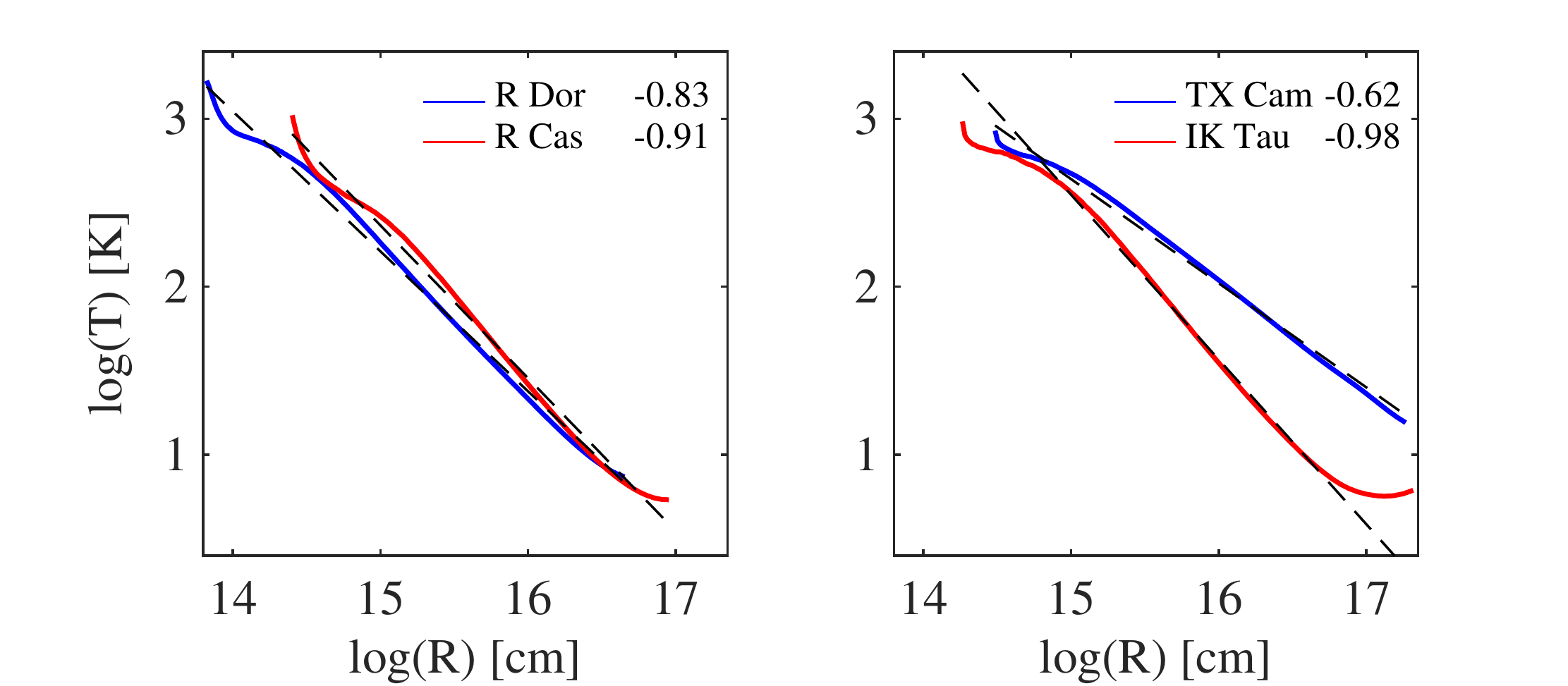}
\caption{Kinetic temperature profiles (solid lines) calculated in the energy balance of the CO modelling for R~Dor (left panel, blue), R~Cas (left panel, red), TX~Cam (right panel, blue), and IK~Tau (right panel, red). The dashed lines show power-law fits to the individual profiles (Eq.~\ref{e:tprofile}). The numbers give the exponents for the respective fits.}
\label{f:tprofiles}
\end{figure}

Finally, the kinetic temperature profiles are shown in Fig.~\ref{f:tprofiles}. Note again that although the energy balance is solved in the CO models, the contribution from \water line cooling is not included in any of the models (see Sect.~\ref{s:heatcool}). For TX~Cam, R~Cas, and R~Dor the profiles can be described well by power laws of the form

\begin{equation}
\label{e:tprofile}
T(r)=T_i\times\left({\frac{r} {R_i}}\right)^\alpha, 
\end{equation}

\noindent
where $T_i$ is the temperature at the inner radius $R_i$. The exponents $\alpha$ are between $-0.6$ and $-1.0$. For the low mass-loss-rate M-type AGB star W~Hya~\cite{khourietal2014a} derive an exponent of $-0.65\pm0.05$. In models of the chemical network in the CSE of AGB stars, \cite{willacyco1997} use $\alpha$ = $-0.6$ for R~Dor, TX~Cam, and IK~Tau. 

Our profile for IK~Tau deviates more from a simple power law in the inner and outermost parts of the CSE, but can generally be fit with $\alpha$ = $-0.98$. Solving the energy balance including \water line cooling in the modelling of CO lines towards IK~Tau, \cite{decinetal2010a} derive a temperature profile consistent with an exponent of $-0.6$, significantly different from our value. Their derived mass-loss rate is $\approx$50\% higher than our value, and they use a dust-to-gas ratio a factor four higher. These values lie within the uncertainties quoted here and in~\cite{decinetal2010a}, and it is not clear why there is a difference in temperature profile and whether this is related to the \water line cooling.

\begin{table*}
\caption{Stellar parameters and results of the SED and CO line-emission models. T$_*$ and L$_*$ are the stellar effective temperature and luminosity, respectively. $R_i$ is the inner radius from which the CSE is modelled, and $T_i$ the temperature at the inner radius. $D$ is the distance to the source. $\tau_{\rm{d}}$ is the dust optical depth at 10\,$\mu$m, \vterminal the terminal gas expansion velocity, $\beta$ the exponent of the velocity law, d/g the derived dust-to-gas ratio, $\dot{M}$ the gas mass-loss rate, $R_{\rm{e}}$ the CO-photodissociation radius, $\delta$ the average ratio between integrated model and observed  line intensities, and $\chi^2_{red}$ the reduced $\chi^2$ value of the best-fit model.}
\label{t:coresults}
\centering
\begin{tabular}{l c c c c c c c c c c c c c}
\hline\hline
Source &  $T_*$	& $L_*$	& $R_i$	& $T_i$	&$D$ & $\tau_d$	& $\upsilon_{\rm{\infty}}$	&$\beta$	&d/g	& $\dot{M}$	& $R_{\rm{e}}$	& $\delta$	& $\chi^2_{red}$\\
			& [K]		& [L$_{\odot}$]&[$10^{13}$\,cm] &[K] & [pc]& & [km/s]	&  &[$10^{-2}$]	&[$10^{-7}$\,M$_{\odot}$]& [$10^{16}$\,cm] & & \\
\hline
R Dor		& 2400	& 6500 & 6.6 &1680& 59	& 0.05	&  5.7	& 1.5		& 0.08	& 1.6& 1.6		& 1.0		& 0.65\\
R Cas	& 1800	& 8725 &25&1050& 172	& 0.09	& 10.5	& 2.5		& 0.04	& 8.0	 & 3.2	&  0.9	& 1.7	\\
TX Cam	& 2400	& 8600 &30&845& 380	& 0.4		& 17.5	& 1.5		& 2.6	& 40 & 6.6			& 0.9		& 3.1\\
IK Tau	& 2100	& 7700 &18&960& 265	& 1.0		&  17.5	& 1.5		& 0.5		& 50 & 7.5	& 0.9		& 2.9	\\
\hline\hline
\end{tabular}
\end{table*}

\subsection{\water models}
\label{s:h2oabundances}

\begin{figure*}[h]
\centering
\includegraphics[width=18cm]{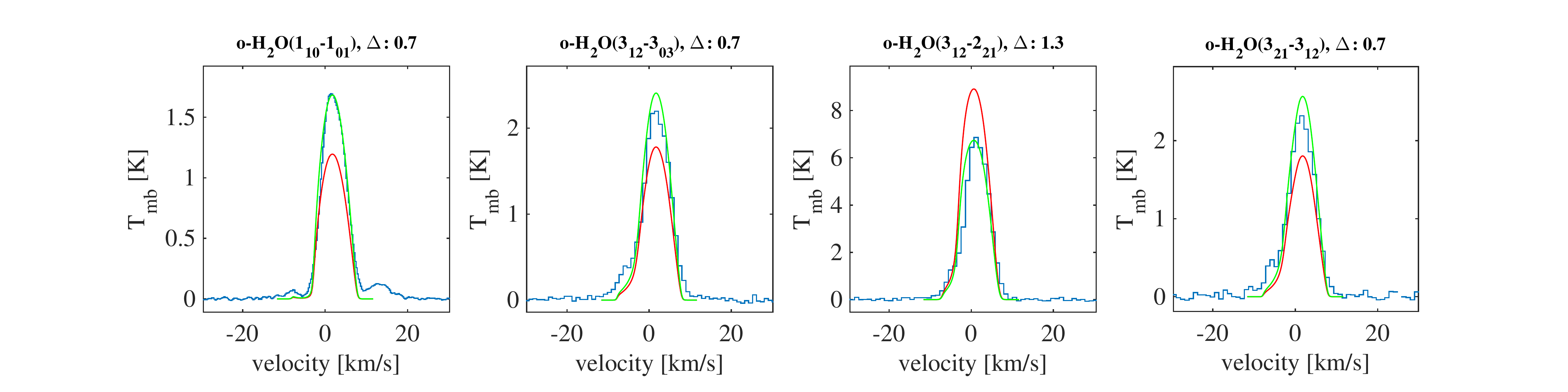}
\includegraphics[width=13.5cm]{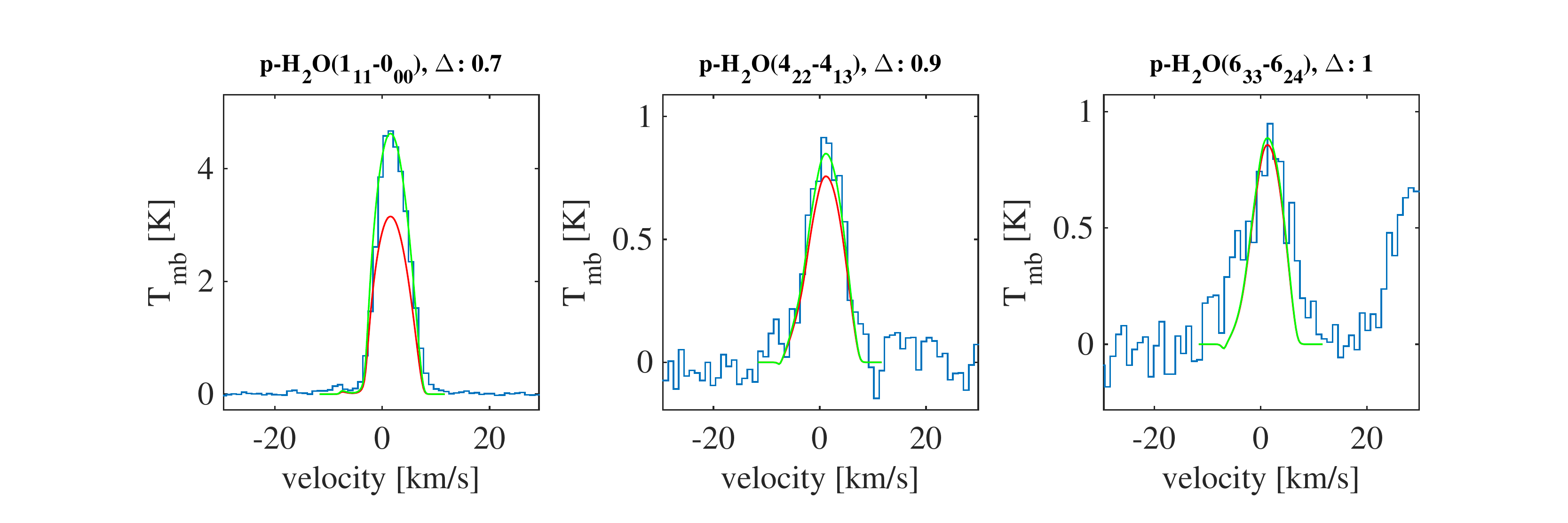}
\caption{Best-fit o-\water and p-\water models of the HIFI Lines for R~Dor. The blue histograms are the observations. The red lines are the model lines, the green lines are the model lines scaled to the same integrated intensities as the observations. The velocities are given with respect to the \vlsr= 6.5\,\kms.}
\label{f:h2omodsrdor}
\end{figure*}

\begin{figure}
\centering
\includegraphics[width=9cm]{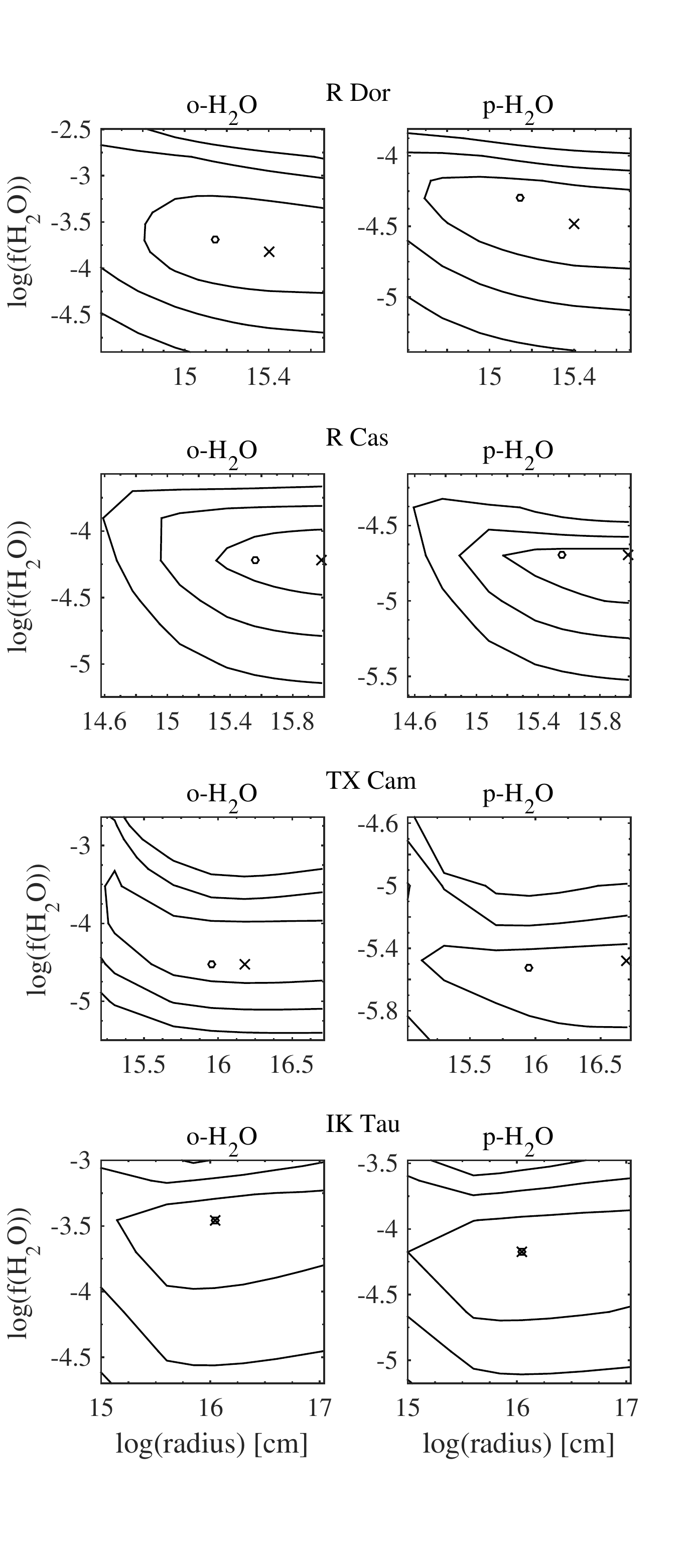}
\caption{$\chi^2$-maps of the o-\water (left) and p-\water (right) model grids for R~Dor , R~Cas , TX~Cam , and IK~Tau  (top to bottom). The cross marks the best-fit model using the \water radius as a free parameter, the circle the best-fit model using R$_{\rm{NK}}$.}
\label{f:h2ochi2maps}
\end{figure}

Figure~\ref{f:h2omodsrdor} shows the best-fit ortho-\water and para-\water models for the HIFI lines compared to the observed lines for R~Dor.  Figures~\ref{f:h2omodsrcas} to~\ref{f:h2omodsiktau} show the equivalent for R~Cas, TX~Cam, and IK~Tau, respectively. Tables~\ref{t:oH2Olines} and~\ref{t:pH2Olines} list the same parameters as Table~\ref{t:COlines}, but for the HIFI and PACS lines of ortho-\water and para-\water, respectively. Figure~\ref{f:h2ochi2maps} shows the $\chi^2_{red}$-maps for ortho-\water and para-\water for all objects. We indicate both the best-fit radius in the grid, as well as the best-fit \water abundance using the radius determined by~\cite{netzerco1987}, \Rnk. It is not possible to set upper limits to the \water radii. The \water line emitting region is limited by radiative excitation, and the \water lines hence do not probe the outer regions very well. In all cases \Rnk is consistent with the best-fit radius when allowing $R$ to vary freely, and we use this radius for the \water envelope. All model results refer to models using \Rnk. Table~\ref{t:h2oresults} gives the best-fit values for the abundance, the average ratios between all integrated model and observed line intensities ($\delta$), and the \chired values. The dependence of the model and observed line ratios on the upper level energy \Eup is shown in Fig.~\ref{f:ratiovsenergy}. As with the CO lines, the scatter lies around one, with a slight under-prediction by the model lines. However, no clear trend with \Eup can be seen.

The derived ortho-\water abundances lie between $\approx$$10^{-5}$--$10^{-4}$ relative to H$_2$. For R~Dor and R~Cas this is significantly lower than previously derived values based only on ISO observations~\citep{maerckeretal2008}. Including the spectrally resolved 557\,GHz line observed with Odin required a reduction of the (constant) expansion velocity to fit the observed line-width, and led to a slight reduction in the ortho-\water abundance for R~Dor, albeit well within the uncertainties~\citep{maerckeretal2009}. Including a velocity profile forces a significant reduction of the \water abundance to reproduce the observed intensities.  Observations of several para-\water lines allow us to set constraints on the para-\water abundances and derive ortho/para-\water ratios (o/p-ratio). The uncertainties in the derived abundances lead to considerable uncertainties in the derived o/p-ratios. This is especially true for TX~Cam, where the low number of observed para-\water lines leads to a bad fit to the observations with the by-far highest \chired value. 

\begin{table*}
\caption{Results of the \water modelling. The \water abundances are given relative to H$_2$. The values in brackets give the 1$\sigma$ ranges.}
\label{t:h2oresults}
\centering
\begin{tabular}{l c c c c c c c c}
\hline\hline
Source 		& $R_{\rm{NK}}$	& f(o-\water)	& $\delta$	& $\chi^2_{red}$ & f(p-\water)	& $\delta$	& $\chi^2_{red}$ & o/p-ratio\\
			& [$\times10^{15}$\,cm]		& [$\times10^{-4}$]&	& &[$\times10^{-4}$]& \\
\hline
R Dor		& 1.4		& 2 (0.7 -- 5.6)		& 0.9 & 1.11 	& 0.5 (0.2 -- 0.7) 	& 1.0 & 1.0	& 4 (1 -- 26)\\
R Cas		& 3.6		& 0.6 (0.4 -- 0.9)	& 0.8 & 1.9	& 0.2 (0.1 -- 0.3) 	& 1.0 & 1.6	& 3 (1 --9)\\
TX Cam		& 9.0		& 0.3 (0.2 -- 1.0)	& 0.6 & 3.9	& 0.03 (0.01 -- 0.04) 	& 0.7 & 6.7	& 10 (5 -- 100)\\
IK Tau		& 11		& 3.5 (1.1 -- 5)		& 0.8 & 1.8	& 0.7 (0.2 -- 1.3) 	& 0.9 & 1.2	& 5 (0.8 -- 25)\\
\hline\hline
\end{tabular}
\end{table*}

\section{Discussion}
\label{s:discussion}

\subsection{The significance of \water line cooling}
\label{s:h2ocool}

The current implementation of \water line cooling in the radiative-transfer codes implies that line cooling by \water significantly affects the temperature in the inner envelope of M-type AGB stars (at radii $R_{\rm{i}}< r <$ a few $R_{\rm{i}}$). The energy balance in this inner part of the CSE is very sensitive to the amount of \water line cooling included in the model, quickly leading to extreme cooling. Models of the S-type stars $\chi$~Cyg and W~Aql, that include \water line cooling, had \water abundances factors $2-35$ lower than the abundances estimated for the M-type stars here~\citep{schoieretal2011,danilovichetal2014}. For our high mass-loss-rate objects, \water line cooling is too strong to be included in the energy balance. For R~Dor and R~Cas we nevertheless managed to include 100\% \water line cooling. While the iterative procedure does not allow us to realistically calculate a grid of models due to computational limitations, the resulting fit to the observed lines gives similar \chired values to the models without \water line cooling. The derived mass-loss rates and \water abundances do not change significantly between models with and without cooling, since the observed CO lines effectively constrain the temperature profile. However, to balance the additional cooling in the inner envelope the d/g ratio needs to be increased by a factor of approximately two. Much more importantly, the only way we managed to include full \water line cooling was to change the drift-velocity profile from starting low and increasing with radius to starting high and decreasing with radius. This changes the dust-velocity profile to a constant value throughout the entire CSE. Although the dust velocity profile and the drift velocity are not constrained observationally, theoretical models of dust-driven winds generally predict very low drift velocities between the dust and the gas for high mass-loss rates, and increasing drift-velocity with increasing distance (as assumed in this paper), or a constant drift velocity. A situation where the dust velocity is constant throughout the envelope therefore seems unphysical. 

A constant dust velocity drastically increases the heating due to the dust-gas interaction in the inner CSE, and hence balances the cooling due to \water. While this scenario is unrealistic, it indicates that some physics leading to heating in the inner envelope may be missing. Since the dust-velocity profile is not constrained by observations, changing the profile in an ad hoc manner becomes similar to adding an arbitrary heating term to counter-balance the cooling by \water. We have no suggestion as to what this additional heating may be, or whether this indicates a fundamental problem in our understanding of the physics in the inner CSEs of AGB stars.

Another possible explanation for the problem of extreme \water line cooling is the way the line cooling is calculated. In the current version of the radiative transfer, the contribution due to line cooling (for any molecule) is included following~\cite{sahai1990}:

\begin{equation}
\label{e:linecool}
C_{line}(r)=\sum_{l,u>l}\Delta E_{ul}k_{B}(p_l c_{lu}-p_u c_{ul}),
\end{equation}

\noindent
where $\Delta E_{ul}$ is the energy difference of a transition from upper level $u$ to lower level $l$, $k_{B}$ the Boltzmann constant, $p_l$ and $p_u$ the lower and upper level population densities, and $c_{lu}$ and $c_{ul}$ the collisional excitation and de-excitation rates from the lower to upper and upper to lower levels, respectively. If the total sum is positive, the rate of downward collisions is lower than the rate of upward collisions. The difference in energy is assumed to be radiated away, leading to net cooling. The contribution to the cooling is calculated in each radial bin. While this approach may be valid for lines that are optically thin throughout the CSE, the fact that no description of the optical depth is included may be problematic at the very high optical depths of the majority of the \water-transitions. The net cooling assumes that the lost energy leaves the radial bin (and CSE). For very high optical depths, however, the radiated energy may be re-absorbed -- either in a radial bin further out, or possibly still in the same radial bin. This would feed energy back into the envelope and reduce the net cooling. The current implementation may hence overestimate the amount of line cooling from \water.

This problem could possibly be solved by simply scaling the contribution to the cooling in Eq.~\ref{e:linecool} by the escape probability $(1-e^{-\tau})/\tau$. The implementation of this in the code is however not trivial. The optical depth depends on the direction and is affected by the surrounding density and velocity profiles. Since the code only works along the radial direction, an angle-averaged optical depth $\bar \tau$ would have to be calculated at each radial point, which is not straight-forward. We are currently working on this problem, trying to address the issue of properly including radiative line cooling into the code.

Note that this does not only affect the cooling due to \water. In principle CO can also reach high optical depths, and in carbon AGB stars cooling due to HCN lines affect the temperature structure in the inner CSE in a very similar way to \water in M-type AGB stars. Including line cooling correctly in the radiative-transfer codes, and identifying any potentially missed heating (or cooling mechanisms) is essential for a complete understanding of the physical conditions in the inner CSEs of AGB stars. This in turn affects chemical and dynamical models of the inner winds, and our understanding of the evolution of the star on the AGB in general.

\subsection{The velocity profiles}
\label{s:vprof}

Although not very well constrained, we derive comparatively shallow velocity profiles for all sources, with $\beta=1.5\pm1.0$. For IK~Tau, \cite{decinetal2010c} derive a $\beta=1-1.8$. The velocity profiles derived here and by \cite{decinetal2010c} are consistent with measurements of maser lines in the CSE of IK~Tau~\citep{bainsetal2003}. The momentum equation for IK~Tau on the other hand gives $\beta=0.6$~\citep{decinetal2010c}. \cite{khourietal2014a} adopt a $\beta=1.5$ to describe the velocity profile in W~Hya, and find evidence that higher values may fit the high-$J$ transitions of CO better. Note that using the upper-state energy $\rm{E_{up}}$ as a measure of the distance in the CSE and the line width to constrain the velocity profile as done by \cite{justtanontetal2012} can be misleading. For R~Dor the measured line width vs. $\rm{E_{up}}$ for all molecules detected in the HIFISTARS observations implies that R~Dor has a decelerating wind~\citep{justtanontetal2012}, showing the importance of using full radiative transfer in order to derive correct velocity profiles.

All measurements indicate that the wind follows a more shallow profile than generally predicted by dynamical models solving the coupled momentum equations of the dust and gas~\citep[e.g.,][]{decinetal2006,ramstedtetal2008,decinetal2010c}. Note however, that \cite{ramstedtetal2008} derive a velocity profile for IK~Tau that is consistent with the results by \cite{bainsetal2003}, showing that it is possible to derive velocity profiles from dynamical models that are consistent with the observed, shallow profiles. At the same time this shows the large uncertainty in the dynamical models, likely due to a sensitive dependence on the input parameters, and the assumption of a fully momentum coupled wind. The choice of $\upsilon_{\rm{g,i}}$ does not significantly affect our results. The observed lines are emitted from too far out in the wind to effectively constrain the wind velocity at the inner radius. Any reasonable changes in $\upsilon_{\rm{g,i}}$ would not change the derived $\beta$ values, but would mainly affect the energy balance in the inner CSE. However, compared to the uncertainty in the treatment of the line cooling, this effect is small.

\subsection{\water abundances}
\label{s:h2oradii}

For all sources we derive total \water abundances between $(0.3--4.0)\times10^{-4}$ relative to \HH (at the inner radius of our model CSE, i.e. $R_i$). Previous modelling of ISO observations of o-\water resulted in the same o-\water abundances for R~Dor and IK~Tau, while the values for R~Cas and TX~Cam were factors 5 and 10 higher, respectively. For IK~Tau, \cite{decinetal2010c} derive a total \water abundance of $6.6\times10^{-5}$, uncertain by a factor of 2. Within our uncertainties we just manage to meet the value derived by \cite{decinetal2010c}. However, they use a mass-loss rate that is higher by $\approx$50\% than our value, likely resulting in a lower \water abundance. For the M-type AGB star W~Hya, previously derived \water abundances are $8\times10^{-4}$ and $3\times10^{-4}$~\citep[at radii smaller and larger than $4.5\times10^{14}$\,cm, respectively;][]{barlowetal1996} ,and $9\times10^{-4}$~\citep{khourietal2014b}. For non-LTE chemical models, the predicted \water abundance in the inner CSE ($<5\,R_*$) of M-type AGB stars is $4\times10^{-4}$ relative to \HH, while LTE models predict \water abundances closer to $7\times10^{-5}$~\citep[using TX~Cam as an example;][]{cherchneff2006}. Non-LTE chemical models of IK~Tau predict \water abundances of $\approx2\times10^{-4}$ at 6\,R$_*$~\citep{gobrechtetal2016}, consistent with our value derived here. Although the \water abundances derived here and in other publications based on radiative-transfer modelling cover a large range, values of a few times $10^{-4}$ appear to be an upper limit. Taken at face value, and assuming that \water abundances remain largely unaffected by processes in the intermediate wind, the results here favour the formation of \water under non-LTE conditions in the inner wind. However, within the uncertainties, the derived abundances seem also to be consistent with the formation of \water under equilibrium conditions at $1\,R_*$. The derived ortho- to para-\water ratio is very uncertain, but is generally consistent with a value of 3, implying formation of \water under warm conditions.

\subsection{Discrepancies between models and observations}
\label{s:discrepancies}

Although the 1D radiative-transfer models generally reproduce the overall line shapes (line-width and double-peaked, flat-topped, parabolic, or triangular lines), it is clear that deviations from these simple profiles are not reproduced. In addition to over-resolved low-$J$ lines in CO, asymmetries between the red- and blue-shifted sides of the lines, and bumps and additional peaks (e.g. in R~Dor and R~Cas, CO($6-5$) at $\approx$3\kms), are not reproduced. In the \water lines the self-absorption on the blue side due to the high optical depths of the lines is generally well reproduced, however also with some deviations. 

These differences may be caused by several effects. Varying mass-loss rates, velocity profiles, and photodissociation efficiencies will affect the spatial density distribution of the circumstellar wind, and the fractional abundance distribution of the molecules. This will change the optical depths of the lines and the excitation conditions. Such effects could, in principle, be included in the 1D radiative transfer. However, effects due to the three dimensional structure may have an equal, or even stronger, effect on the observed lines. These include asymmetric winds, disks and tori, and clumpiness of the circumstellar matter. Shocks in the inner wind and asymmetric illumination of the CSE due to convection cells on the stellar surface and atmospheric molecular absorption bands can also strongly affect the observed line shapes. Since the beginning of science observations with ALMA (Atacama Large Millimeter/submillimeter Array), an increasingly complex picture of the inner winds of AGB stars is being established, including the effects mentioned above. VLT-SPHERE observations in the optical of R~Dor resolve the stellar disk, and show clear signs of an irregular brightness distribution from the stellar surface due to convection cells (Khouri et al. in preparation).

\subsection{Comparison to radiative transfer with GASTRoNOoM}
\label{s:codecomp}

Solving the molecular line radiative transfer for CSEs around AGB stars encounters several physical (e.g. energy balance, radiation fields, velocity profiles, molecular data) and numerical (e.g. modelling strategies, matrix inversion, optical depths)  problems that have to be solved. As such, the number of advanced, non-LTE radiative-transfer codes in use is limited. This is unfortunate, since this makes benchmarking, and comparison and verification of model results difficult. One of the radiative-transfer codes that approaches the modelling of AGB CSEs in a similar way to our code is GASTRoNOoM~\citep{decinetal2006,decinetal2007,decinetal2008a}.

The estimate of the mass-loss rate for IK~Tau based on GASTRoNOoM lies within 50\% of our results~\citep{decinetal2010c}.  \cite{khourietal2014a} derive a mass-loss rate for W~Hya of $1.5\times10^{-7}$\,\Msunyr, compared to $1.0\times10^{-7}$\,\Msunyr in our previous modelling~\citep{maerckeretal2008}. Based on a grid of models using GASTRoNOoM, \cite{debecketal2010} derive a formula to estimate the mass-loss rates from AGB stars based on the intensities of the CO lines. Using their formula results in mass-loss rates that are less than a factor two different from the values derived here. This is well within the limit of the absolute uncertainty expected from this modelling method~\citep{ramstedtetal2008}. Compared to what we find in our modelling, the derived \water abundances using GASTRoNOoM appear to be lower by a factor of two for IK~Tau. Comparing the results by \cite{khourietal2014b} with \cite{maerckeretal2008}, they are a factor of five lower for W~Hya. However, the absolute uncertainties are relatively large, and a more systematic comparison between the two radiative-transfer codes is necessary to identify whether the discrepancy in \water abundances is due to numerical differences in the codes, or simply reflects the uncertainty and difficulty in modelling \water line emission.

\section{Conclusions}
\label{s:conclusions}

We have successfully modelled CO and \water lines observed towards four M-type AGB stars, including spectrally resolved ground-based and HIFI observations for CO, spectrally resolved HIFI observations of o-\water and p-\water, and spectrally unresolved observations of CO and \water with PACS. These are the first models of a larger sample of sources with spectrally resolved \water line profiles. We derive \water abundances, o/p-\water ratios, the sizes of the \water line emitting regions, velocity and temperature profiles, and attempt to include full \water line cooling in the energy balance in the radiative transfer. Based on the modelling we arrive at following conclusions.

\begin{itemize}
\item Our modelling indicates that the winds around these four M-type AGB stars have a comparatively shallow acceleration profile, in contrast to what is generally expected from dynamical models of dust-driven winds. However, we also note that the widths of the line profiles only moderately constrain the velocity profile.
\item Although we managed to include \water line cooling in the energy balance for the low mass-loss rate objects R~Dor and R~Cas, this required an unphysical description of the dust expansion velocity in order to counter-balance the extreme cooling from \water in the inner CSE. This implies that the physical processes that describe the heating and cooling in the inner envelope are not completely understood and/or that the formal implementation of line cooling is not correct. 
\item We derive generally high \water abundances, up to a few times $10^{-4}$ relative to \HH. Within the uncertainties, all derived abundances are consistent with both NLTE and LTE abundances derived from chemical models, emphasizing the importance of improved radiative transfer models to reduce uncertainties on abundances derived from observations.
\item The derived o/p-ratios are uncertain, however consistent with a value of three, indicating formation of \water in the warm, inner CSE.
\end{itemize}

Computational limitations currently do not allow us to self-consistently solve the physics and dynamics throughout the CSE in three dimensions. 1D radiative-transfer models as presented here are important for constraining the basic physical parameters of the envelopes (e.g., temperature, velocity, average mass-loss rates). These can be used as input for 3D models ~\citep[e.g. LIME;][]{brinchco2010} to study the effect of structure in three dimensions. In order to arrive at a self-consistent model that fully describes the physical conditions throughout the CSE, however, it is important to solve the problem of radiative line cooling, in particular for \water.

\begin{acknowledgements}
M.M. has received funding from the People Programme (Marie Curie Actions) of the EU's FP7 (FP7/2007-2013) under REA grant agreement No. 623898.11. HO acknowledges financial support from the Swedish Research Council. M.M. would like to thank Fredrik Sch\"oier for his contribution at the beginning of this project, his advice and support. Sadly he could not be here to see the results.
\end{acknowledgements}

\bibliographystyle{aa} 
\bibliography{maercker}

\appendix

\section{Observed and model line intensities, and model line profiles}
\label{s:Aobsmodint}

\begin{table*}
\caption{Observed and modelled CO lines.}
\label{t:COlines}
\centering
\tiny
\begin{tabular}{l cccccccccccccccccc}
\hline\hline
 & &&  &\multicolumn{3}{c}{R Dor} & & \multicolumn{3}{c}{R Cas} & & \multicolumn{3}{c}{TX Cam} & & \multicolumn{3}{c}{IK Tau}  \\
Transition& $\lambda$&E$_{\mathrm{up}}$&  &I$_{\mathrm{obs}}$& I$_{\mathrm{mod}}$& $\Delta$&  &I$_{\mathrm{obs}}$& I$_{\mathrm{mod}}$& $\Delta$&  &I$_{\mathrm{obs}}$& I$_{\mathrm{mod}}$& $\Delta$&  &I$_{\mathrm{obs}}$& I$_{\mathrm{mod}}$& $\Delta	$	\\
$J$-level & [GHz] &[K] & & \multicolumn{3}{c}{[K \kms]}  &  & \multicolumn{3}{c}{[K \kms]}  &  & \multicolumn{3}{c}{[K \kms]}  &  &\multicolumn{3}{c}{[K \kms]}  \\
\hline
\textbf{Ground based}			 &				&	&			&			&		&	&			&			&		&	&			&			&		&	&			&			&		\\
CO$(1-0)$				&115.3&	3.85		&&5.00    	&5.52    	&1.10	&	&8.40   &8.21    	&0.98	&&14.5		&13.0			&0.90	&&	33.0	&	32.0	&	0.97	\\
CO$(2-1)$				&230.5&	11.54		&&41.6		&38.5		&0.93	&	&32.1	&35.1		&1.09	&&60.6		&43.7			&0.72	&&	95.0	&	74.1	&	0.78	\\
CO$(3-2)$				&345.8&	23.07		&&70.4		&72.4		&1.03	&	&100	&68.3		&0.68	&&70.0		&87.4			&1.25	&&	125		&	110		&	0.88	\\
CO$(4-3)$				&461.0&	38.45		&&--		&	--		&--		&	&89.6	&85.7		&0.96	&&149		&95.7			&0.64	&&	130		&	130		&	1.00	\\
						 &&				&	&			&				&		&	&			&			&		&	&			&			&		&	&			&			&		\\
$J$-level & [GHz] & [K] & & \multicolumn{3}{c}{[K \kms]}  &  & \multicolumn{3}{c}{[K \kms]}  &  & \multicolumn{3}{c}{[K \kms]}  &  &\multicolumn{3}{c}{[K \kms]}  \\
\hline
\textbf{HIFI}			 &&				&	&			&			&		&	&			&			&		&	&			&			&		&	&			&			&		\\
CO$(6-5)	$			&691.5&80.74	&	&16.8	&	12.5	&	0.74&	&17.2	&	7.74    &	0.45&	&16.6	&	8.19   &	0.49&			&12.0	&10.9	&0.91\\	
CO$(10-9)	$			&1152.0&211.40	&	&15.9	&	13.8	&	0.87&	&12.8	&	8.87    &	0.69&	&8.15   	&9.86   &	1.21&			&9.45	&13.7	&1.45	\\
CO$(16-15)	$			&1841.3&522.48	&	&11.6	&	13.8	&	1.19&	&6.76	&	7.69    &	1.14&	&--		&--	&--		&					&15.9	&14.1	&0.89	\\
						 &&				&	&			&			&		&	&			&			&		&	&			&			&		&	&			&			&		\\
$J$-level & [$\mu$m]& [K] & &\multicolumn{3}{c}{[$10^{-16}$\,W\,m$^{-2}$]}  &  & \multicolumn{3}{c}{[$10^{-16}$\,W\,m$^{-2}$]}  &  & \multicolumn{3}{c}{[$10^{-16}$\,W\,m$^{-2}$]}  &  &\multicolumn{3}{c}{[$10^{-16}$\,W\,m$^{-2}$]}  \\
\hline
\textbf{PACS}			 &				&	&			&			&		&	&			&			&		&	&			&			&		&	&			&			&		\\
CO$(14-13)$				&186.00&403.46	&&	2.43    &	2.78    &	1.14&	&1.45    	&1.65    	&1.14&	&2.06    &	1.96    &	0.95&	&3.63	&2.85	&0.79\\	
CO$(15-14)$				&173.63&461.05	&&	2.94    &	2.96    &	1.01&	&1.53    	&1.70    	&1.11&	&2.07    &	2.06    &	1.00&	&4.55	&3.03	&0.67\\	
CO$(16-15)$				&162.81&522.48	&&	--		&	--		&	--	&	&1.70    	&1.74    	&1.02&	&2.07    &	2.14    &	1.03&	&3.93	&3.18	&0.81\\	
CO$(17-16)$				&153.27&587.72	&&	3.77    &	3.26    &	0.86&	&1.70    	&1.75    	&1.03&	&1.93    &	2.20    &	1.14&	&4.06	&3.30	&0.81\\	
CO$(19-18)$				&137.20&729.68	&&	--		&	--		&	--	&	&1.55    	&1.72    	&1.11&	&--		&	--		&	--	&	&4.06	&3.44	&0.85\\	
CO$(21-20)$				&124.19&886.90	&&	3.77    &	3.55    &	0.94&	&1.35    	&1.63    	&1.21&	&8.77    &	2.21    &	0.25&	&3.53	&3.46	&0.98\\	
CO$(22-21)$				&118.58&971.23	&&	--		&	--		&	--	&	&3.16    	&1.57    	&0.50&	&3.18    &	2.17    &	0.68&	&5.79	&3.42	&0.59\\	
CO$(24-23)$				&108.76&1151.32&&		--	&		--	&	--	&	&--			&--			&--	&	&1.40    &	2.03    &	1.45&	&4.16	&3.27	&0.79\\	
\hline\hline
\end{tabular}
\end{table*}

\newpage

\begin{table*}
\caption{Observed and modelled ortho-\water lines.}
\label{t:oH2Olines}
\centering
\tiny
\begin{tabular}{l cccccccccccccccccc}
\hline\hline
 & &&  &\multicolumn{3}{c}{R Dor} & & \multicolumn{3}{c}{R Cas} & & \multicolumn{3}{c}{TX Cam} & & \multicolumn{3}{c}{IK Tau}  \\
Transition& $\lambda$&E$_{\mathrm{up}}$&  &I$_{\mathrm{obs}}$& I$_{\mathrm{mod}}$& $\Delta$&  &I$_{\mathrm{obs}}$& I$_{\mathrm{mod}}$& $\Delta$&  &I$_{\mathrm{obs}}$& I$_{\mathrm{mod}}$& $\Delta$&  &I$_{\mathrm{obs}}$& I$_{\mathrm{mod}}$& $\Delta	$	\\
$J_{\mathrm{K}_\mathrm{a}, \mathrm{K}_\mathrm{c}}$-level & [GHz]& [K] & & \multicolumn{3}{c}{[K \kms]}  &  & \multicolumn{3}{c}{[K \kms]}  &  & \multicolumn{3}{c}{[K \kms]}  &  &\multicolumn{3}{c}{[K \kms]}  \\
\hline
\textbf{HIFI}			 &&				&	&			&			&		&	&			&			&		&	&			&			&		&	&			&			&		\\
o-\water$(1_{10}-1_{01})$&556.9&	60.96		&	&12.3		&8.72		&0.71	&	&8.36		&3.87		&0.46	&	&6.71		&2.45		&0.36	&	&11.4	&6.8 	&0.59	\\
o-\water$(5_{32}-4_{41})$&620.7&	732.06		&	&--			&--			&--		&	&--			&--			&--		&	&--			&--			&--		&	&18.2	&28.2	&1.55	\\	
o-\water$(3_{12}-3_{03})$&1097.4&	249.43		&	&18.6		&13.7		&0.74	&	&13.1		&6.36		&0.49	&	&7.24		&4.68		&0.65	&	&15.7	&14.1	&0.89	\\
o-\water$(3_{12}-2_{21})$&1153.1&	249.43		&	&51.6		&68.3		&1.32	&	&24.9		&23.1		&0.93	&	&10.6		&13.2		&1.24	&	&28.4	&34.4	&1.21	\\
o-\water$(3_{21}-3_{12})$&1162.9&	305.24		&	&19.3		&13.6		&0.70	&	&10.4		&5.02		&0.48	&	&4.91		&3.07		&0.62	&	&15.1	&11.2	&0.74	\\
o-\water$(3_{02}-2_{12})$&1716.8&	196.77		&	&--			&--			&--		&	&--			&--			&--		&	&--			&--			&--		&	&32.2	&30.1	&0.93	\\
						 &				&	&			&			&		&	&			&			&		&	&			&			&		&	&			&			&		\\
$J_{\mathrm{K}_\mathrm{a}, \mathrm{K}_\mathrm{c}}$-level & [$\mu$m]&[K] & &\multicolumn{3}{c}{[$10^{-16}$\,W\,m$^{-2}$]}  &  & \multicolumn{3}{c}{[$10^{-16}$\,W\,m$^{-2}$]}  &  & \multicolumn{3}{c}{[$10^{-16}$\,W\,m$^{-2}$]}  &  &\multicolumn{3}{c}{[$10^{-16}$\,W\,m$^{-2}$]}  \\
\hline
\textbf{PACS}			 &&				&	&			&			&		&	&			&			&		&	&			&			&		&	&			&			&		\\
o-\water$(2_{21}-2_{12})$&180.49&	194.09		&	&5.92		&5.51		&0.93	&	&2.65		&1.25		&0.47	&	&3.77		&1.00		&0.27	&	&5.93	&2.14	&0.36	\\
o-\water$(2_{12}-1_{01})$&179.53&	114.38		&	&14.3		&14.1		&0.99	&	&7.32		&3.39		&0.46	&	&8.56		&2.56		&0.30	&	&10.0	&4.18	&0.42	\\
o-\water$(7_{34}-7_{25})$&166.81&	1211.96		&	&2.32		&1.35		&0.58	&	&0.33		&0.32		&0.97	&	&--			&--			&--		&	&2.52	&1.31	&0.52	\\
o-\water$(5_{32}-5_{23})$&160.51&	732.06		&	&4.27		&3.16		&0.74	&	&1.07		&0.86		&0.81	&	&0.79		&0.63		&0.80	&	&4.20	&2.21	&0.53	\\
o-\water$(8_{45}-7_{52})$&159.05&	1615.32		&	&4.87		&7.54		&1.55	&	&0.56		&0.61		&1.08	&	&--			&--			&--		&	&1.94	&2.25	&1.16	\\
o-\water$(5_{14}-5_{05})$&134.94&	574.73		&	&6.48		&6.38		&0.99	&	&1.71		&1.70		&1.00	&	&--			&--			&--		&	&6.37	&4.65	&0.73	\\
o-\water$(8_{36}-7_{43})$&133.55&	1447.57		&	&10.1		&18.2		&1.80	&	&--			&--			&--		&	&--			&--			&--		&	&4.45	&5.44	&1.22	\\
o-\water$(4_{23}-4_{14})$&132.41&	432.15		&	&9.88		&8.10		&0.82	&	&3.28		&2.01		&0.61	&	&2.51		&1.58		&0.63	&	&9.47	&4.59	&0.48	\\
o-\water$(9_{45}-9_{36})$&129.34&	1957.06		&	&--			&--			&--		&	&--			&--			&--		&	&--			&--			&--		&	&2.61	&1.56	&0.60	\\	
o-\water$(7_{25}-7_{16})$&127.88&	1125.71		&	&3.07		&2.35		&0.77	&	&--			&--			&--		&	&--			&--			&--		&	&5.82	&3.16	&0.54	\\
o-\water$(9_{36}-9_{27})$&123.46&	1845.82		&	&2.72		&1.29		&0.47	&	&--			&--			&--		&	&--			&--			&--		&	&2.23	&1.75	&0.79	\\
o-\water$(4_{32}-4_{23})$&121.72&	550.35		&	&9.33		&6.65		&0.71	&	&3.17		&1.61		&0.51	&	&1.60		&1.34		&0.84	&	&9.86	&5.00	&0.51	\\
o-\water$(7_{34}-6_{43})$&116.78&	1211.96		&	&15.7		&16.7		&1.06	&	&--			&--			&--		&	&--			&--			&--		&	&7.48	&6.56	&0.88	\\
o-\water$(2_{21}-1_{10})$&108.07&	194.09		&	&32.7		&27.8		&0.85	&	&13.3		&6.76		&0.51	&	&10.2		&4.94		&0.48	&	&24.5	&10.2	&0.41	\\
o-\water$(6_{34}-6_{25})$&104.09&	933.73		&	&5.96		&5.43		&0.91	&	&--			&--			&--		&	&--			&--			&--		&	&10.8	&6.23	&0.58	\\
o-\water$(6_{25}-6_{16})$&94.64&	795.51		&	&11.3		&8.99		&0.80	&	&3.05		&2.52		&0.83	&	&--			&--			&--		&	&12.4	&8.22	&0.66	\\
o-\water$(6_{43}-6_{34})$&92.81&	1088.75		&	&8.60		&6.68		&0.78	&	&1.60		&1.72		&1.07	&	&--			&--			&--		&	&10.3	&7.40	&0.72	\\
o-\water$(8_{36}-8_{27})$&82.98&	1447.57		&	&5.74		&5.42		&0.94	&	&1.16		&1.44		&1.24	&	&--			&--			&--		&	&9.33	&8.08	&0.87	\\
o-\water$(6_{16}-5_{05})$&82.03&	643.49		&	&--			&--			&--		&	&9.97		&9.85		&0.99	&	&--			&--			&--		&	&9.44	&6.35	&0.67	\\
o-\water$(9_{27}-9_{18})$&81.41&	1729.29		&	&--			&--			&--		&	&1.65		&1.04		&0.63	&	&--			&--			&--		&	&29.8	&17.4	&0.58	\\
o-\water$(4_{23}-3_{12})$&78.74&	432.15		&	&50.4		&37.1		&0.74	&	&16.3		&11.0		&0.68	&	&--			&--			&--		&	&7.19	&8.07	&1.12	\\
o-\water$(7_{52}-7_{43})$&77.76&	1524.86		&	&6.16		&5.42		&0.88	&	&--			&--			&--		&	&--			&--			&--		&	&11.4	&10.4	&0.91	\\
o-\water$(5_{50}-5_{41})$&75.91&	1067.68		&	&13.7		&9.65		&0.70	&	&1.99		&2.16		&1.09	&	&--			&--			&--		&	&36.6	&22.6	&0.62	\\
o-\water$(3_{21}-2_{12})$&75.38&	305.24		&	&51.4		&57.2		&1.11	&	&20.9		&13.2		&0.63	&	&--			&--			&--		&	&18.0	&18.5	&1.03	\\
o-\water$(7_{07}-6_{16})$&71.95&	843.47		&	&30.4		&36.9		&1.21	&	&--			&--			&--		&	&--			&--			&--		&	&14.7	&12.9	&0.88	\\
o-\water$(8_{27}-8_{18})$&70.70&	1274.17		&	&19.0		&11.4		&0.60	&	&--			&--			&--		&	&--			&--			&--		&	&--			&--			&--		\\
o-\water$(3_{30}-3_{03})$&67.27&	410.65		&	&51.9		&32.0		&0.62	&	&13.8		&7.05		&0.51	&	&--			&--			&--		&	&--			&--			&--		\\
o-\water$(3_{30}-2_{21})$&66.44&	410.65		&	&62.8		&50.6		&0.81	&	&17.8		&11.6		&0.65	&	&--			&--			&--		&	&--			&--			&--		\\
o-\water$(7_{16}-6_{25})$&66.09&	1013.20		&	&42.1		&48.6		&1.16	&	&--			&--			&--		&	&--			&--			&--		&	&--			&--			&--		\\
o-\water$(6_{25}-5_{14})$&65.17&	795.51		&	&40.7		&36.3		&0.89	&	&--			&--			&--		&	&--			&--			&--		&	&--			&--			&--		\\
o-\water$(8_{18}-7_{07})$&63.32&	1070.68		&	&52.0		&42.9		&0.83	&	&--			&--			&--		&	&--			&--			&--		&	&--			&--			&--		\\
o-\water$(4_{32}-3_{21})$&58.70&	550.35		&	&54.3		&46.5		&0.86	&	&14.2		&13.1		&0.92	&	&--			&--			&--		&	&--			&--			&--		\\
\hline\hline
\end{tabular}
\end{table*}

\begin{table*}
\caption{Observed and modelled para-\water lines.}
\label{t:pH2Olines}
\centering
\tiny
\begin{tabular}{l cccccccccccccccccc}
\hline\hline
 & && & \multicolumn{3}{c}{R Dor} & & \multicolumn{3}{c}{R Cas} & & \multicolumn{3}{c}{TX Cam} & & \multicolumn{3}{c}{IK Tau}  \\
Transition& $\lambda$ &E$_{\mathrm{up}}$&  &I$_{\mathrm{obs}}$& I$_{\mathrm{mod}}$& $\Delta$&  &I$_{\mathrm{obs}}$& I$_{\mathrm{mod}}$& $\Delta$&  &I$_{\mathrm{obs}}$& I$_{\mathrm{mod}}$& $\Delta$&  &I$_{\mathrm{obs}}$& I$_{\mathrm{mod}}$& $\Delta	$	\\
$J_{\mathrm{K}_\mathrm{a}, \mathrm{K}_\mathrm{c}}$-level & [GHz]&[K] & & \multicolumn{3}{c}{[K \kms]}  &  & \multicolumn{3}{c}{[K \kms]}  &  & \multicolumn{3}{c}{[K \kms]}  &  &\multicolumn{3}{c}{[K \kms]}  \\
\hline
\textbf{HIFI}			 &				&	&			&			&		&	&			&			&		&	&			&			&		&	&			&			&		\\
p-\water$(2_{11}-2_{02})$	&752.0&136.94	&&	--			&--			&--		&&--		&	--		&	--	&&	--		&	--		&	--	&&	13.2	&10.9	&0.82	\\
p-\water$(5_{24}-4_{31})$	&970.3&598.83	&&	--			&--			&--		&&--		&	--		&	--	&&	--		&	--		&	--	&&	17.5	&33.7	&1.93	\\
p-\water$(2_{02}-1_{11})$	&987.9&100.84	&&	--			&--			&--		&&--		&	--		&	--	&&	--		&	--		&	--	&&	23.7	&17.0	&0.72	\\
p-\water$(1_{11}-0_{00})$	&1113.3&53.54	&&	33.6		&22.9		&0.68	&&19.8		&10.4		&0.53	&&13.2		&5.31		&0.40	&&	24.4	&16.5	&0.68\\
p-\water$(4_{22}-4_{13})$	&1207.6&454.33	&&	6.72		&5.99		&0.89	&&3.36		&1.98		&0.59	&&	--		&	--		&	--	&&	--			&--			&--	\\
p-\water$(5_{33}-6_{06})$	&1717.0&725.09	&&	--			&--			&--		&&--		&	--		&	--	&&	--		&	--		&	--	&&	2.70	&3.01	&1.11\\		
p-\water$(6_{33}-6_{24})$	&1762.0&951.82	&&	6.18		&5.97		&0.97	&&--		&	--		&	--	&&	--		&	--		&	--	&&	2.10	&7.49	&3.57\\
						 &				&	&			&			&		&	&			&			&		&	&			&			&		&	&			&			&		\\
$J_{\mathrm{K}_\mathrm{a}, \mathrm{K}_\mathrm{c}}$-level & [$\mu$m]&[K] & &\multicolumn{3}{c}{[$10^{-16}$\,W\,m$^{-2}$]}  &  & \multicolumn{3}{c}{[$10^{-16}$\,W\,m$^{-2}$]}  &  & \multicolumn{3}{c}{[$10^{-16}$\,W\,m$^{-2}$]}  &  &\multicolumn{3}{c}{[$10^{-16}$\,W\,m$^{-2}$]}  \\
\hline
\textbf{PACS}			 &				&	&			&			&		&	&			&			&		&	&			&			&		&	&			&			&		\\
p-\water$(4_{13}-4_{04})$&187.11		&396.38		&	&2.61    	&2.66    	&1.02	&&0.74    	&0.95    	&1.29	&&0.94    	&0.35    	&0.37	&&2.28	&2.17	&0.95	\\
p-\water$(7_{35}-6_{42})$&169.74		&1175.03		&	&2.74    	&4.05    	&1.48	&&0.53    	&0.58    	&1.10	&&--		&	--		&	--	&&1.66	&3.21	&1.93	\\
p-\water$(6_{24}-6_{15})$&167.03		&867.25		&	&1.08    	&1.05    	&0.97	&&--		&	--		&	--	&&	--		&	--		&	--	&&2.64	&1.72	&0.65	\\
p-\water$(4_{31}-4_{22})$&146.92		&552.26		&	&3.34    	&3.66    	&1.10	&&1.05    	&1.09    	&1.04	&&0.82    	&0.34    	&0.42	&&4.20	&2.93	&0.70	\\
p-\water$(4_{13}-3_{22})$&144.52		&396.38		&	&11.6    	&19.8    	&1.71	&&3.22    	&4.29    	&1.33	&&2.63    	&1.11    	&0.42	&&8.46	&6.62	&0.78	\\
p-\water$(4_{04}-3_{13})$&125.35		&319.48		&	&20.2    	&25.9    	&1.28	&&--		&	--		&	--	&&0.88    	&1.78    	&2.03	&&15.0	&7.59	&0.51	\\
p-\water$(9_{37}-8_{44})$&118.41		&1749.88		&	&3.30    	&1.86    	&0.56	&&--		&	--		&	--	&&	--		&	--		&	--	&&1.83	&2.51	&1.37	\\
p-\water$(9_{46}-8_{53})$&117.68		&1929.22		&	&--			&--			&--		&&--		&	--		&	--	&&	--		&	--		&	--	&&1.53	&1.95	&1.27	\\
p-\water$(5_{24}-5_{15})$&111.63		&598.83		&	&6.18    	&5.82    	&0.94	&&1.92    	&1.77    	&0.92	&&0.97    	&0.68    	&0.70	&&7.28	&6.12	&0.84	\\
p-\water$(5_{42}-5_{33})$&94.21		&877.81		&	&7.38    	&5.01    	&0.68	&&1.49    	&1.29    	&0.87	&&--		&	--		&	--	&&	--		&	--		&	--	\\
p-\water$(7_{44}-7_{35})$&90.05		&1334.81		&	&--			&--			&--		&&--		&	--		&	--	&&	--		&	--		&	--	&&5.29	&5.91	&1.12	\\
p-\water$(6_{06}-5_{15})$&83.28		&642.69		&	&31.9    	&31.7    	&0.99	&&5.81    	&7.33    	&1.26	&&--		&	--		&	--	&&20.2	&14.6	&0.72\\
p-\water$(8_{35}-7_{44})$&81.69		&1510.93		&	&--			&--			&--		&&--		&	--		&	--	&&	--		&	--		&	--	&&4.79	&82.4	&1.72	\\	
p-\water$(7_{26}-7_{17})$&81.22		&1020.93		&	&--			&--			&--		&&--		&	--		&	--	&&	--		&	--		&	--	&&9.42	&9.68	&1.03	\\
p-\water$(8_{53}-8_{44})$&80.56		&1806.97		&	&--			&--			&--		&&--		&	--		&	--	&&	--		&	--		&	--	&&4.60	&5.76	&1.25	\\
p-\water$(6_{15}-5_{24})$&78.93		&781.12		&	&29.1    	&31.5    	&1.08	&&--		&	--		&	--	&&	--		&	--		&	--	&&19.8	&16.7	&0.84\\
p-\water$(6_{51}-6_{42})$&76.42		&1278.54		&	&--		&	--		&	--		&&--		&	--		&	--	&&	--		&	--		&	--	&&7.63	&9.08	&1.19	\\
p-\water$(9_{37}-9_{28})$&73.61		&1749.88		&	&--		&	--		&	--		&&	--		&	--		&	--	&&	--		&	--		&	--	&&5.54	&6.74	&1.22	\\
p-\water$(5_{51}-6_{24})$&71.79		&1067.67		&	&--		&	--		&	--		&&--		&	--		&	--	&&	--		&	--		&	--	&&6.20	&0.62	&0.10		\\
p-\water$(7_{17}-6_{06})$&71.54		&843.81		&	&25.3    	&34.1    	&1.35	&&--		&	--		&	--	&&	--		&	--		&	--	&&15.1	&17.0	&1.13	\\
p-\water$(5_{24}-4_{13})$&71.07		&598.83		&	&29.4    	&35.4    	&1.20	&&6.91    	&8.91    	&1.29	&&--		&	--		&	--	&&	\\
p-\water$(3_{31}-2_{20})$&67.09		&410.36		&	&50.8    	&35.9    	&0.71	&&--		&	--		&	--	&&	--		&	--		&	--	&&--		&	--		&	--	\\
p-\water$(8_{08}-7_{17})$&63.46		&1070.54		&	&24.8    	&31.5    	&1.27	&&--		&	--		&	--	&&	--		&	--		&	--	&&--		&	--		&	--	\\
p-\water$(4_{31}-4_{04})$&61.81		&552.26		&	&25.8    	&19.1    	&0.74	&&--		&	--		&	--	&&	--		&	--		&	--	&&--		&	--		&	--	\\
p-\water$(8_{26}-7_{35})$&60.16		&1749.82		&	&19.3    	&22.0    	&1.14	&&--		&	--		&	--	&&	--		&	--		&	--	&&--		&	--		&	--	\\
p-\water$(7_{26}-6_{15})$&59.99		&1414.18		&	&31.7    	&30.5    	&0.96	&&--		&	--		&	--	&&	--		&	--		&	--	&&--		&	--		&	--	\\
\hline\hline
\end{tabular}
\end{table*}

\newpage

\begin{figure*}[h]
\centering
\vspace{-0.5cm}
\includegraphics[width=19cm]{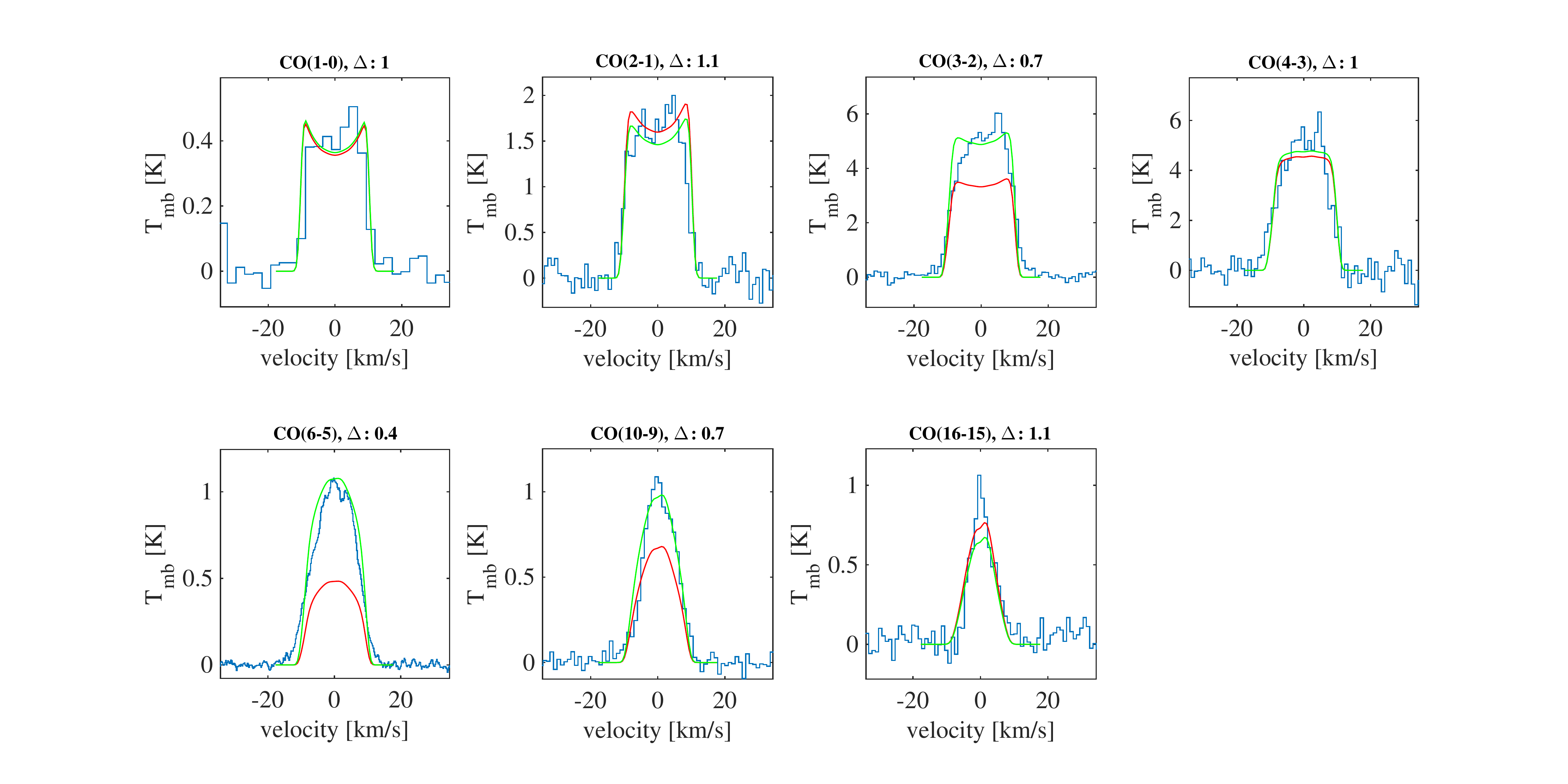}
\caption{Best-fit CO models for R~Cas. The blue histograms are the observations. The red lines are the model lines, the green lines are the model lines scaled to the same integrated intensities as the observations. The velocities are given with respect to the \vlsr= 24.5\,\kms.}
\label{f:colinesrcas}
\end{figure*}

\begin{figure*}
\centering
\includegraphics[width=19cm]{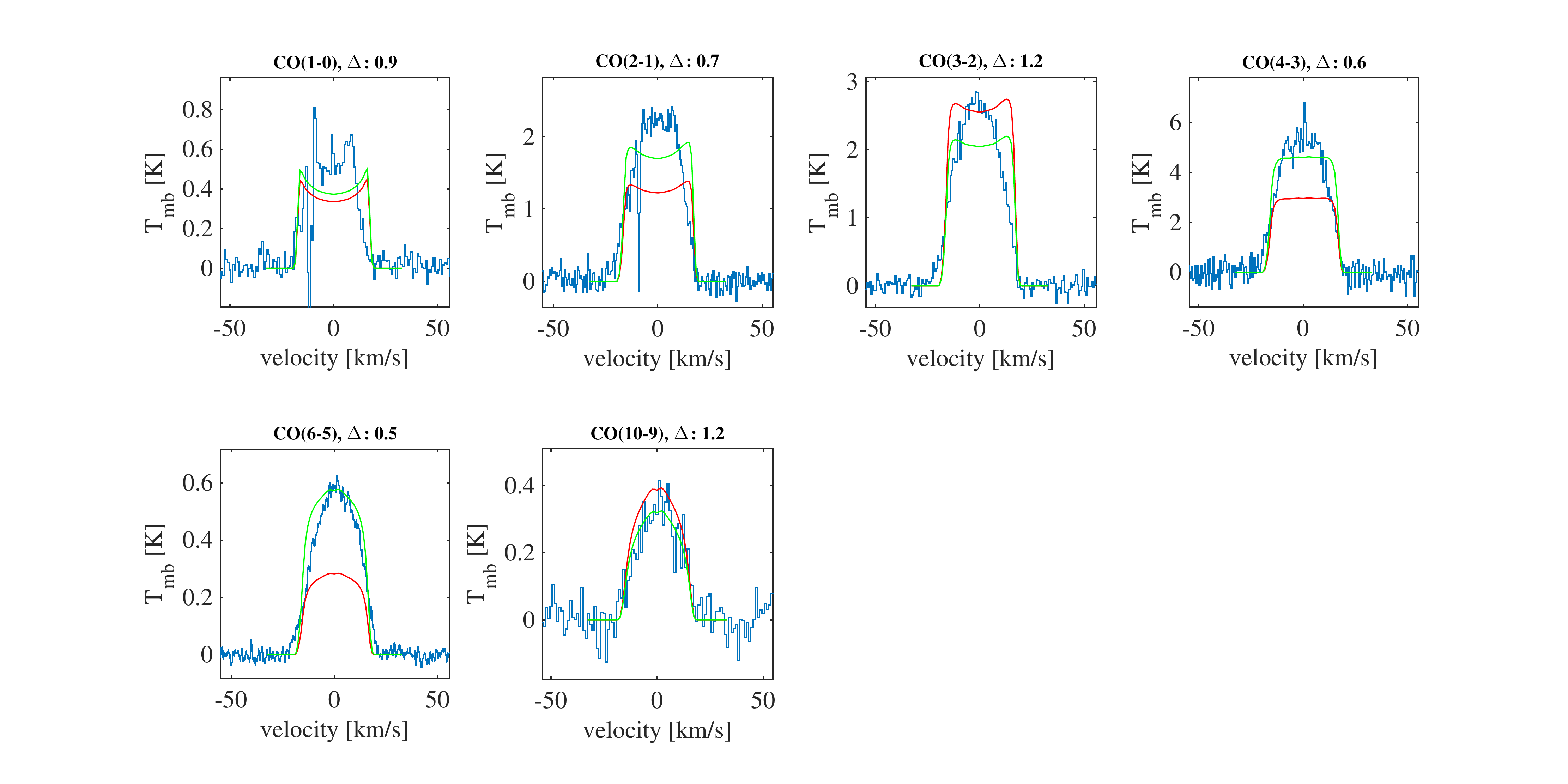}
\caption{Best-fit CO models for TX~Cam.The blue histograms are the observations. The red lines are the model lines, the green lines are the model lines scaled to the same integrated intensities as the observations. The velocities are given with respect to the \vlsr= 12.0\,\kms.}
\label{f:colinestxcam}
\end{figure*}

\begin{figure*}
\centering
\includegraphics[width=19cm]{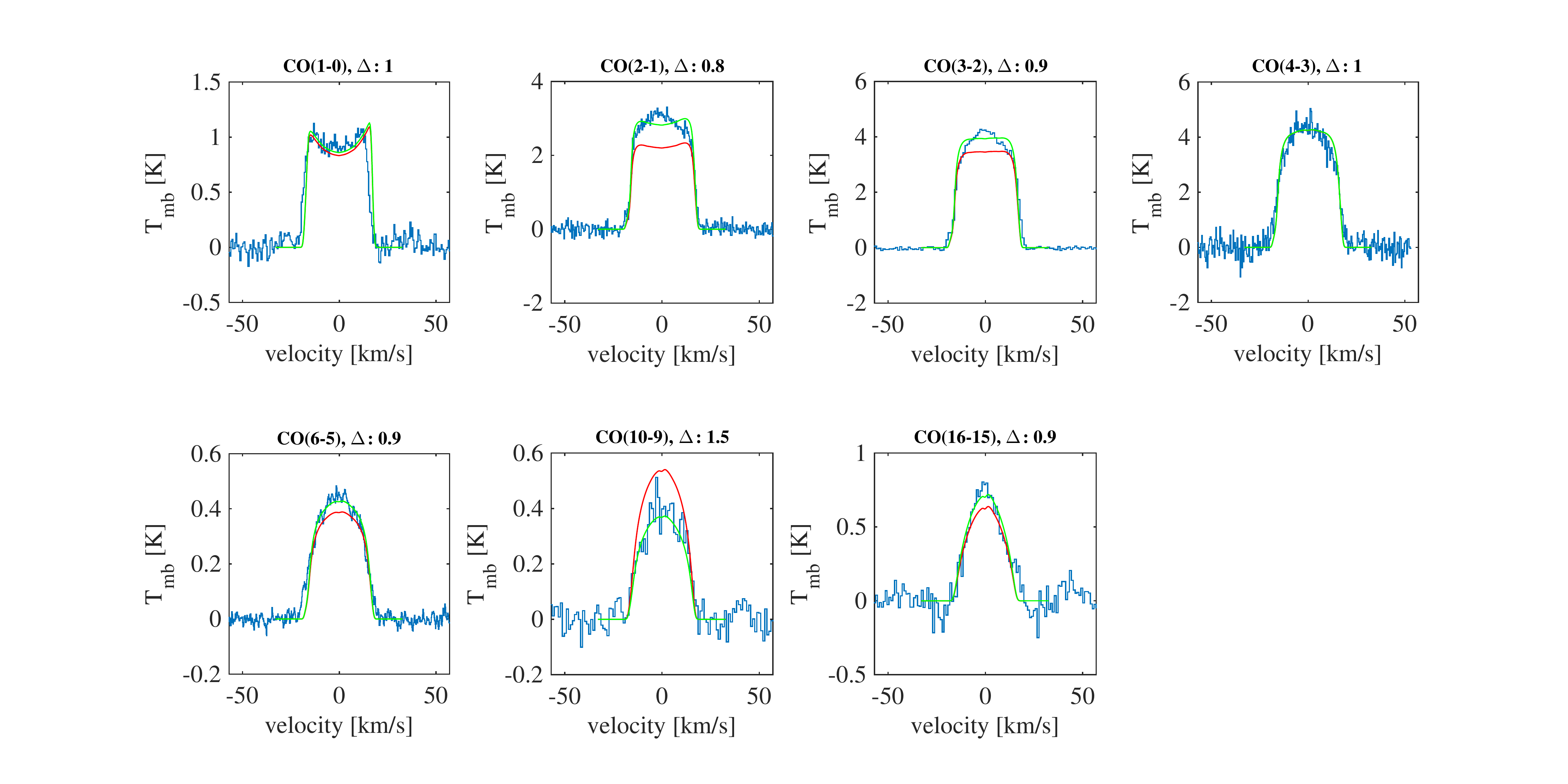}
\caption{Best-fit CO models for IK~Tau.The blue histograms are the observations. The red lines are the model lines, the green lines are the model lines scaled to the same integrated intensities as the observations. The velocities are given with respect to the \vlsr= 34.0\,\kms.}
\label{f:colinesiktau}
\end{figure*}

\begin{figure*}
\centering
\includegraphics[width=18cm]{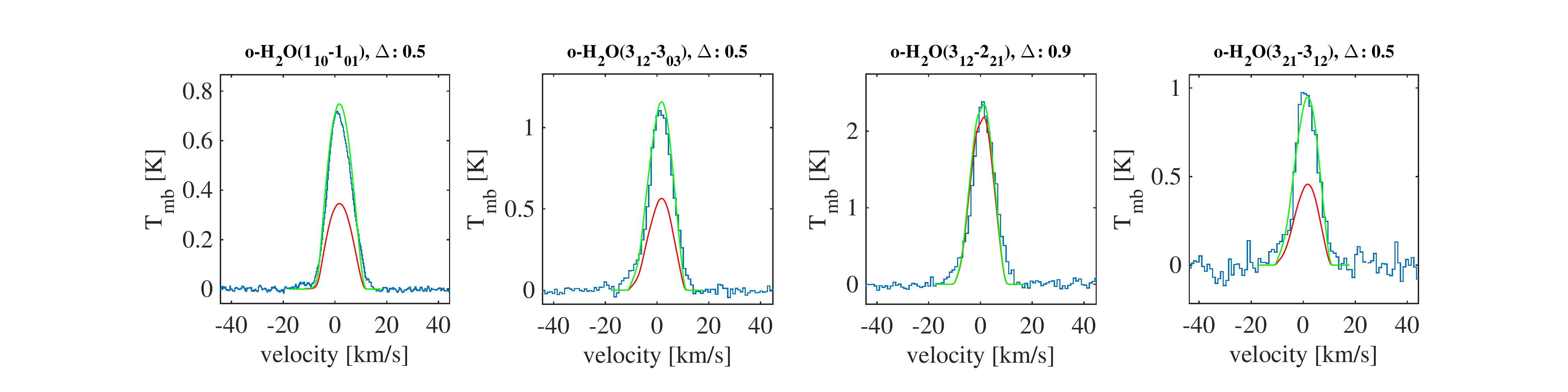}
\includegraphics[width=9cm]{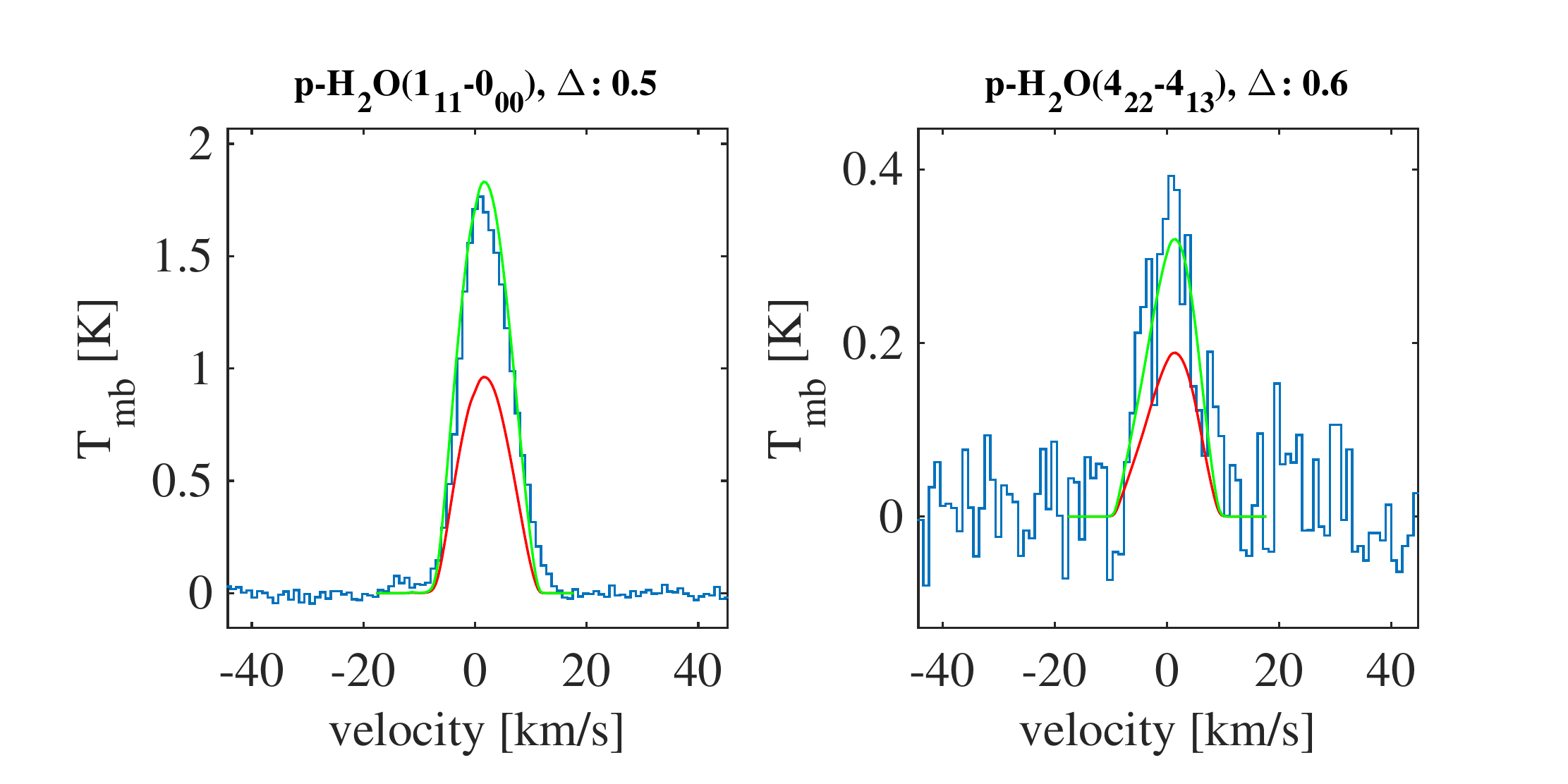}
\caption{Best-fit o-\water and p-\water models of the HIFI Lines for R~Cas. The blue histograms are the observations. The red lines are the model lines, the green lines are the model lines scaled to the same integrated intensities as the observations. The velocities are given with respect to the \vlsr= 24.5\,\kms.}
\label{f:h2omodsrcas}
\end{figure*}

\begin{figure*}
\centering
\includegraphics[width=18cm]{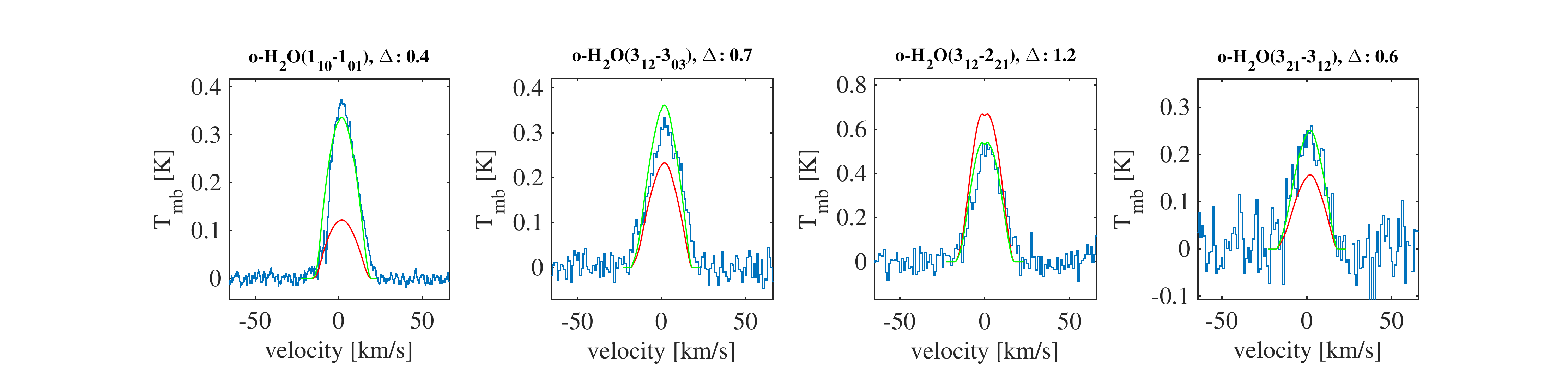}
\includegraphics[width=4.cm]{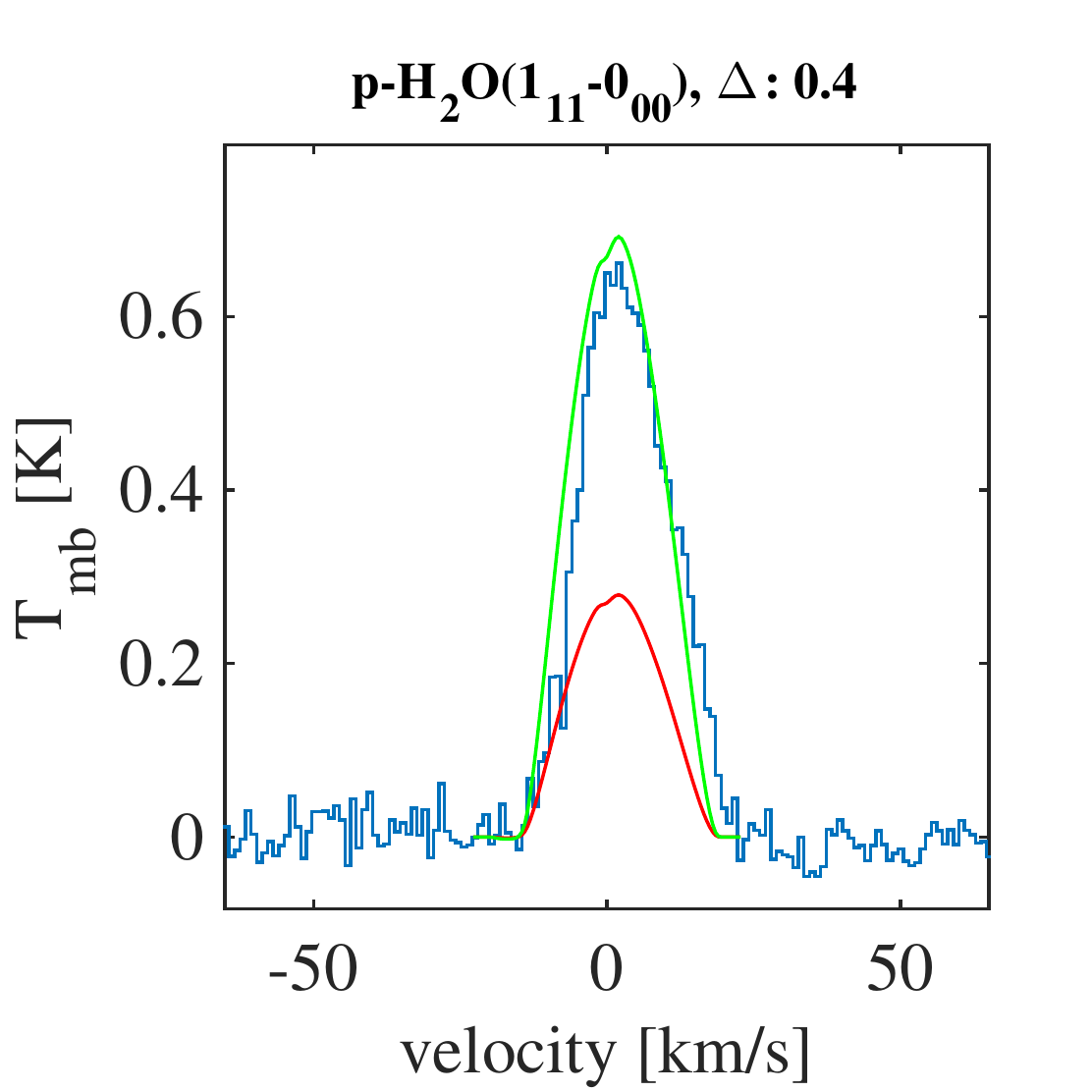}
\caption{Best-fit o-\water and p-\water models of the HIFI Lines for TX~Cam. The blue histograms are the observations. The red lines are the model lines, the green lines are the model lines scaled to the same integrated intensities as the observations. The velocities are given with respect to the \vlsr= 12.0\,\kms.}
\label{f:h2omodstxcam}
\end{figure*}

\begin{figure*}
\centering
\includegraphics[width=18cm]{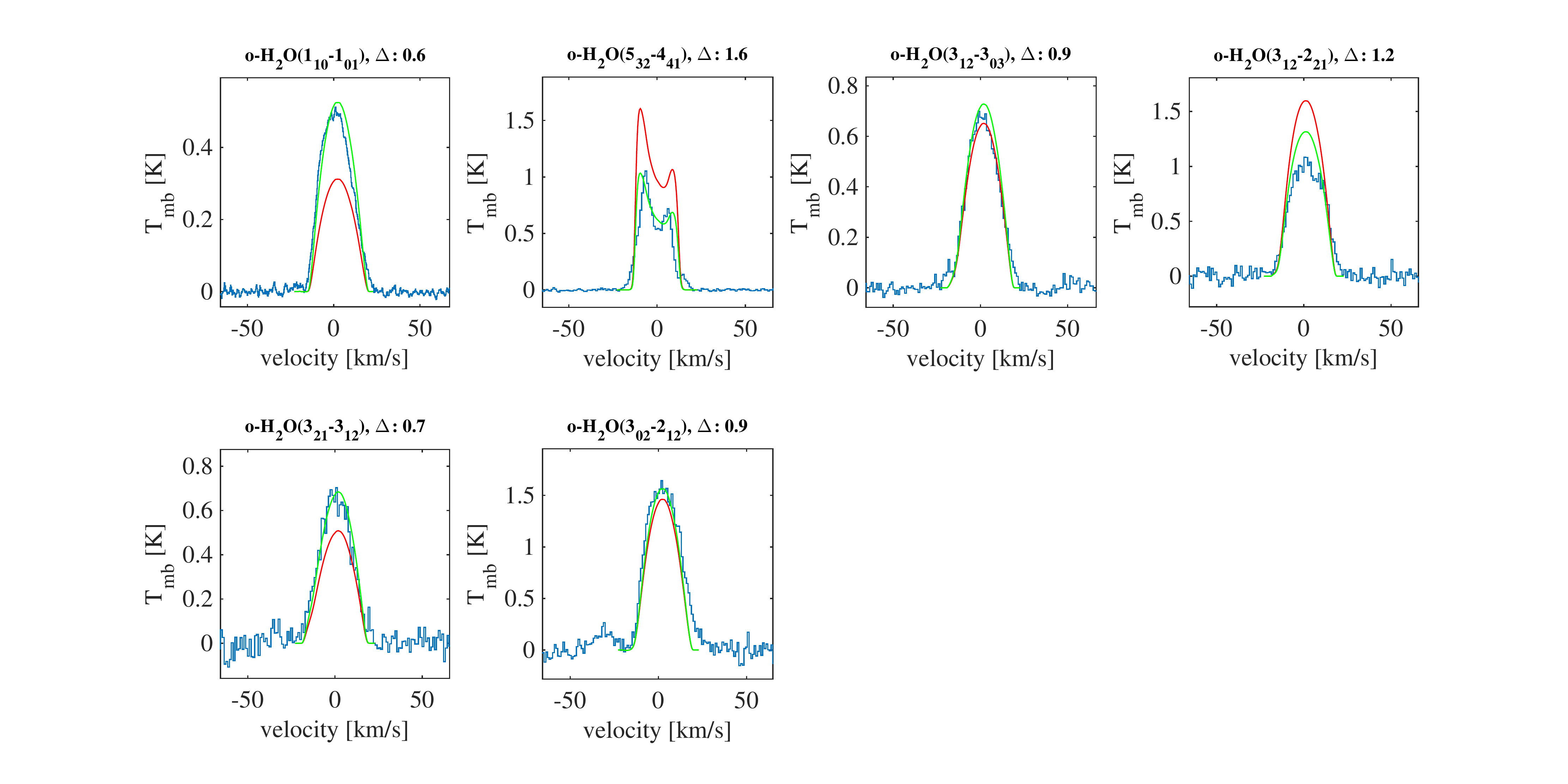}
\includegraphics[width=18cm]{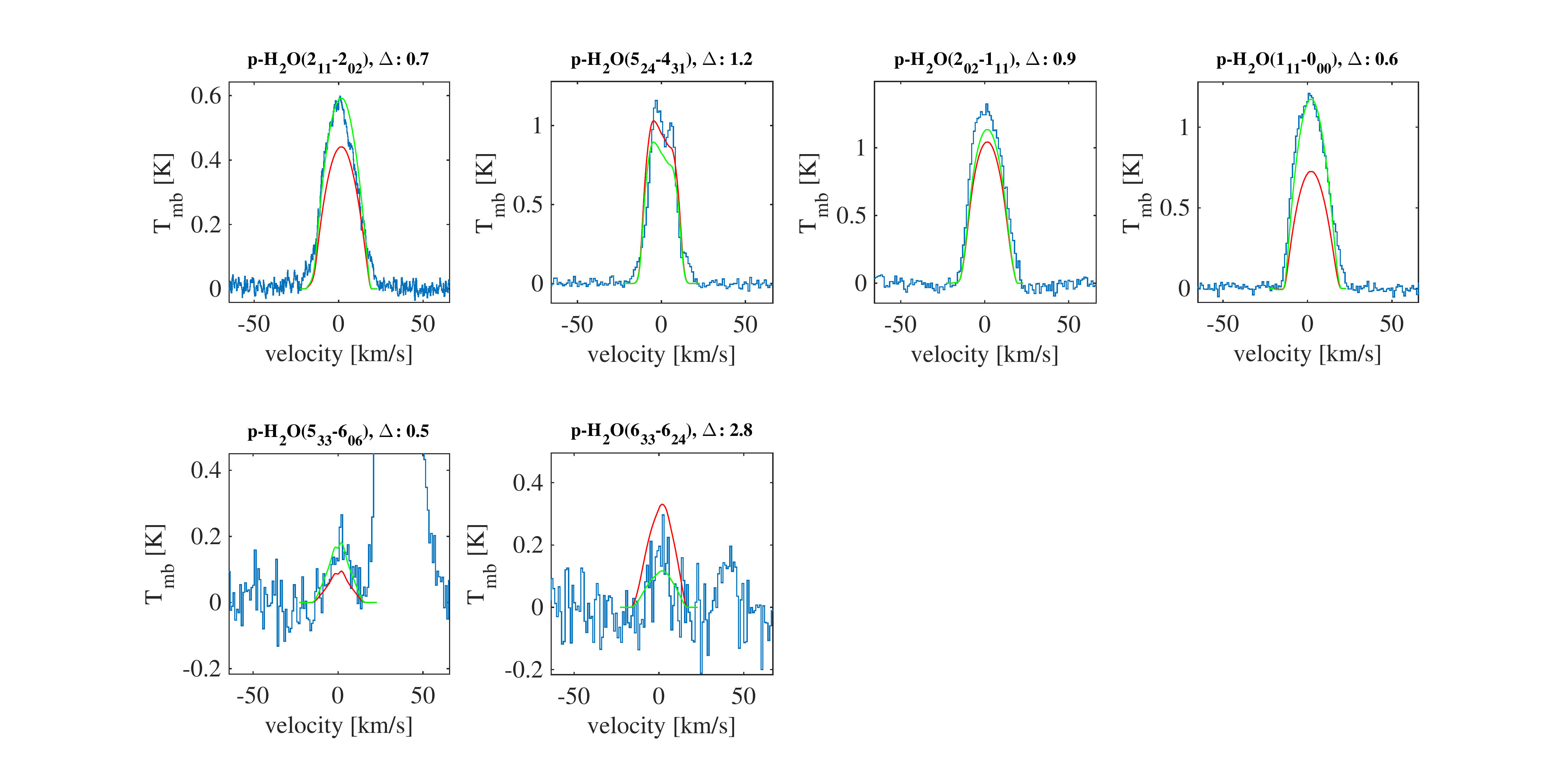}
\caption{Best-fit o-\water and p-\water models of the HIFI Lines for IK~Tau. The blue histograms are the observations. The red lines are the model lines, the green lines are the model lines scaled to the same integrated intensities as the observations. The velocities are given with respect to the \vlsr= 34.0\,\kms.}
\label{f:h2omodsiktau}
\end{figure*}

\end{document}